\documentclass[%
 reprint,
superscriptaddress,
 amsmath,amssymb,
 aps,
pre,
]{revtex4-2}

\usepackage{graphicx}% Include figure files
\usepackage{dcolumn}% Align table columns on decimal point
\usepackage{bm}% bold math
\usepackage{url} % for URLs in bibliography
\begin{document}

% \preprint{APS/123-QED}

\title{Characterization of Phase Transitions with the Annealed Importance Sampling Algorithm}% Force line breaks with \\

\author{A. Prat Pou}
\email{arnau.prat.pou@upc.edu}
 \affiliation{Departament de F\'isica, Universitat Polit\`ecnica de Catalunya, Barcelona Tech, Campus Nord B4-B5, E-08034 Barcelona, Spain.}%Lines break automatically or can be forced with \\

\author{E. Romero}%
 \email{eromero@cs.upc.edu}
\affiliation{Departament de Ci\`encies de la Computaci\'o, Universitat Polit\`ecnica de Catalunya, Barcelona Tech, Campus Nord B4-B5, E-08034, Barcelona, Spain }%

% \collaboration{MUSO Collaboration}%\noaffiliation

\author{J. Mart\'i}
 \email{jordi.marti@upc.edu}
\affiliation{Departament de F\'isica, Universitat Polit\`ecnica de Catalunya, Barcelona Tech, Campus Nord B4-B5, E-08034 Barcelona, Spain.}%
% \affiliation{
%  Third institution, the second for Charlie Author
% }%
\author{F. Mazzanti}
 \email{ferran.mazzanti@upc.edu}
\affiliation{Departament de F\'isica, Universitat Polit\`ecnica de Catalunya, Barcelona Tech, Campus Nord B4-B5, E-08034 Barcelona, Spain.}%

% \collaboration{CLEO Collaboration}%\noaffiliation


\begin{abstract}
We explore the use of the Annealed Importance Sampling (AIS) algorithm as a numerical tool to study phase transitions and critical phenomena. By mapping physical systems such as Ising and spin-glass models directly onto Restricted Boltzmann Machines (RBMs), and extending the analysis to both synthetic and trained architectures, we use a unified RBM framework to compare the performance of AIS against standard Monte Carlo methods in regimes characterized by strong fluctuations and complex energy landscapes. The unified RBM formulation allows for massively parallel sampling, offering a scalable and precise alternative to traditional Monte Carlo methods. Building on a simple, heuristic mean-field AIS initialization derived from the weights of the network, the proposed scheme enhances sampling efficiency and accuracy across all systems considered, proving its effectiveness as a tool to detect and characterize phase transitions in discrete spin and spin-like systems. Our results show that a properly initialized AIS reliably estimates thermodynamic observables, including the partition function and magnetization, while accurately detecting both paramagnetic–ferromagnetic and paramagnetic–spin-glass phase transitions across various system sizes.

\end{abstract}

\maketitle

%\tableofcontents

%%%%%%%%%%%%%%%%%%%%%%%%%%%%%%%%%%%%%%%%%%%%%%%%%%%%%%%%%%%%%%%%%%%%%%%%%%%
\section{Introduction}
The study of phase transitions and critical phenomena has long been a central topic in statistical physics, as it provides a fundamental framework to understand the bridge between macroscopic and microscopic properties. Through the calculation of the partition function $Z$, one gains access to key thermodynamic quantities such as the free energy, entropy, and response functions, which characterize the global behavior of a system~\cite{Intro_PT, Intro_SP}. From a formal point of view, phase transitions can be associated with discontinuities in the free energy or its derivative, according to the Ehrenfest classification~\cite{PT_Classification1}, or with the emergence (or disappearance) of an order parameter that captures the qualitative distinction between phases, according to the Landau classification~\cite{PT_Classification2,Intro_PT}.

Well-known examples include the liquid--gas transition in fluids~\cite{Transition_water}, the paramagnetic--ferromagnetic transition in magnetic materials~\cite{Onsager}, the onset of superconductivity~\cite{Transition_BSC}, the formation of a Bose--Einstein condensate~\cite{Transition_BEC}, and order–disorder transitions in binary alloys~\cite{Transition_BinaryAlloy}. Despite their very different microscopic origins, many of these systems display remarkably similar behavior near their transition points. This observation led to the concept of universality, according to which systems with different microscopic details can share the same critical behavior, characterized by universal critical exponents and scaling laws~\cite{Universality1,Universality2}.

The region close to the critical point is particularly challenging to study. In this regime, thermal fluctuations dominate the statistical properties of the system, correlation lengths diverge, and mean-field or perturbative approaches often fail. Physical observables exhibit strong fluctuations near the transition point, diverging in the thermodynamic limit for continuous transitions or showing finite discontinuities and latent heat in the case of first-order phase transitions~\cite{PT_Classification2,Intro_PT}. These features make analytical treatments difficult and, in many cases, intractable, especially for interacting systems of high dimensionally or with intrinsic disorder such as in magnetic systems (Ising, Sherrington-Kirkpatrick, Edwards-Anderson, some of them discussed below), for many-fermion systems (Sachdev-Ye-Kitaev model~\cite{Heisenberg_magnet}) or in Hopfield neural networks~\cite{Hopfield}. Near the critical point, probability distributions become highly nontrivial, metastable states may coexist, and the configuration space can develop large free-energy barriers separating relevant regions. Under these conditions, simple local-update algorithms often fail to efficiently explore phase space, leading to long autocorrelation times and poor statistical estimates. As a result, numerical approaches play a crucial role in the investigation of critical phenomena.

Monte Carlo methods have therefore become one of the most powerful tools for studying phase transitions in complex systems~\cite{MC_SP1, MC_SP2}. However, near critical points, standard algorithms often suffer from critical slowing down, motivating the development of more sophisticated sampling strategies. Several advanced Monte Carlo algorithms have been proposed to address these limitations. Some examples are Parallel Tempering~\cite{Parallel_Tempering}, Wolff algorithm~\cite{Wolff}, Swendsen--Wang algorithm~\cite{Swendsen_Wang}, and Wang--Landau algorithm~\cite{WangLandau}.

In this work, we focus on AIS~\cite{AIS1,AIS2}, an algorithm that simulates an annealing process from a simple probability distribution $p_0(\mathbf{x})$ to a target distribution $p(\mathbf{x})$. AIS enables the estimation of the partition function $Z$ as well as expectation values of observables $\langle \mathcal{O} \rangle$. Neal showed that AIS converges to the exact values of $Z$ and $\langle \mathcal{O} \rangle$ in the limit of an infinite number of samples and annealing steps. In practice, however, such limits are unattainable, and the performance of the algorithm critically depends on the choice of parameters, mainly the amount of statistics available and, last but not least, the initialization scheme.

Although AIS has been applied to several physical systems (see, for example,~\cite{AIS_Quimica, AIS_MD, AIS_Biologia}) and to the RBM partition function estimation~\cite{AIS_ZRBM, Salak_08}, it remains relatively underused in the context of phase transitions and critical phenomena. The main goal of this work is to demonstrate that AIS is not only a powerful method for estimating thermodynamic quantities --- particularly at low temperatures --- but also a reliable tool for detecting phase transitions. We compare the performance of AIS with commonly used algorithms such as Metropolis--Hastings (MH), Gibbs Sampling (GS), and Parallel Tempering (PT), highlighting the observed enhancement of performance offered by AIS in direct comparison to those approaches, including in the RBM setting~\cite{RBM_PTtraining1, RBM_PTtraining2}. Furthermore, we systematically validate the AIS initialization scheme proposed in~\cite{Article_original} across a broad class of spin-like and neural-network models, demonstrating for the first time that it reliably detects and characterizes phase transitions, even under stringent computational constraints. While several refinements of AIS have been proposed in the literature~\cite{AIS_Millora_PalantePatras, AIS_Millora_CR, AIS_Millora_LIS, AIS_Millora_SMC, AIS_Millora_SW, AIS_diff}, none has focused on the detailed numerical detection of phase transition phenomena, to the best of our knowledge.

We test AIS on several spin-like systems, including the Ising model~\cite{Ising}, two spin glass systems~\cite{Edwards_1975,BipartiteSK_model}, and RBMs~\cite{RBM_1}. All these systems can be mapped onto a bilayer network structure, which allows for efficient parallelization of the simulations using GPUs, following an approach similar to that of~\cite{mAIS}. We show that, with an appropriate initialization, AIS accurately samples relevant thermodynamic quantities such as the partition function $Z$ and the magnetization $m$, and successfully identifies the different phase transitions present in these systems.

The article is organized as follows. In Sec.~\ref{Sec:RBM}, we introduce RBMs and the bilayer representation used in our simulations. In Sec.~\ref{Sec:AIS}, we describe the AIS algorithm and detail the initialization scheme proposed in \cite{Article_original}. Section~\ref{Sec:Results} presents our results: Sec.~\ref{Subsec:Results_AISB0} focuses on Ising and spin glass systems, and proves that AIS is able to obtain good estimations of critical parameters of said models, while Sec.~\ref{Subsec:Results_AISSgh} addresses artificial and trained RBMs, and proves the advantages of using the mentioned AIS initialization scheme against standard AIS schemes.

\section{Restricted Boltzmann Machines}
\label{Sec:RBM}
An RBM is an energy-based neural network composed of a \textit{visible layer} and a \textit{hidden layer}~\cite{RBM_1,RBM_2,Smolensky1986}, and such that interactions are restricted to units in different layers. Despite their relatively simple structure, RBMs can represent intricate correlations between variables~\cite{HintonSalakhutdinov2006,Bengio_09}. Typical RBM models use binary units in the $\{0,1\}$ or $\{-1,1\}$ range, and have an associated energy

\begin{equation}
    E(\mathbf{x}) = - \mathbf{v}^T \mathbf{b} - \mathbf{h}^T 
    \mathbf{c} - \mathbf{v}^T \mathbf{W} \mathbf{h},
    \label{eq:RBM_ene}
\end{equation}
where $\mathbf{v}, \mathbf{h}$ represents the visible and hidden configurations, $\mathbf{W}$ is the two-body interaction matrix and $\mathbf{b}, \mathbf{c}$ are the bias of the visible and hidden units.

The joint probability distribution of visible and hidden units follows the Boltzmann form
\begin{equation}
    p(\mathbf{v}, \mathbf{h}) = \frac{e^{-E(\mathbf{v}, \mathbf{h})/T}}{Z},
    \label{eq:RBM_prob}
\end{equation}
where $Z$ is the partition function, $T$ is the temperature, and we set $k_B = 1$. The partition function is defined as
\begin{equation}
    Z = \sum_{\{\mathbf{v}, \mathbf{h}\}}e^{-E(\mathbf{v}, \mathbf{h})/T}~,
\end{equation}
where the sum extends though all of the possible $2^{N_v+N_h}$ visible and hidden configurations, being $N_v, N_h$ the total number of visible and hidden units. Due to the bipartite structure, it is always possible to sum over one of the two components (visible or hidden) units. For units in $\{0,1\}$, summing over the hidden layer reduces to
\begin{equation}
    Z = \sum_{\{\mathbf{v}\}} e^{\mathbf{v}^T \mathbf{b}/T} \prod_{j=1}^{N_h}     \left(1 + e^{(c_j + \mathbf{v}^T \mathbf{W}_j) / T} \right),
    \label{eq:RBM_Z}
\end{equation}
where $\mathbf{W}_j$ denotes the $j$-th column of $\mathbf{W}$. By symmetry, one can equivalently sum over the hidden units.

Another magnitude of interest is the magnetization. For a given configuration $ [\mathbf{v}, \mathbf{h}]$, the magnetization is defined as
\begin{equation}
    \label{Eq:magne}
    m(\mathbf{v}, \mathbf{h}) = \frac{1}{N}\left[ \sum_{i = 1}^{N_v}v_i + \sum_{j = 1}^{N_h}h_j \right]~,
\end{equation}
where $N = N_v + N_h$ is the total number of units. Similarly to the partition function, the exact mean magnetization can be evaluated exactly. For the $\{0, 1\}$ case, the equation reads as
\begin{multline}
    \langle m \rangle = \frac{1}{Z N} \sum_{\{\mathbf{v}\}} 
    e^{\mathbf{v}^T \mathbf{b}/T} 
    \left( \sum_{i=1}^{N_v} v_i + \sum_{j=1}^{N_h} \sigma_j \right) \\
    \times\prod_{j=1}^{N_h} \left( 1 + e^{(\mathbf{v}^T \mathbf{W}_j + c_j)/T} \right),
    \label{Eq:magne_RBM}
\end{multline}
where $\sigma_j \equiv \sigma\!\left((\mathbf{v}^T \mathbf{W}_j + c_j)/T\right)$ is the sigmoid function $\sigma(x) = 1/\left(1 + e^{-x}\right)$. The derivation can be found in the supplementary material section, together with the derivation for the $\{-1, 1\}$ case.

Restricted Boltzmann Machines can also map Ising-like systems with nearest-neighbor interactions onto a bilayer structure~\cite{Weigel_1, Weigel_2, Weigel_3}, which facilitates parallelization and allows the direct application of the sampling and partition function estimation methods discussed in Sec.~\ref{Sec:AIS}, similarly to what was done in~\cite{Article_original}. All systems analyzed in this work were therefore mapped into an RBM.

\section{Annealed Importance Sampling}
\label{Sec:AIS}
Annealed Importance Sampling (AIS) is a Monte Carlo algorithm that enables sampling from a very complex or directly intractable probability distribution $p(\mathbf{x})$ allowing also to estimate the partition function $Z$ and any statistical average $\langle O\rangle_{p({\bf x})} = \int d{\bf x}\, p({\bf x}) O({\bf x})$~\cite{AIS1,AIS2}. The algorithm proceeds by gradually transforming a tractable initial distribution $p_0(\mathbf{x})$ into $p(\mathbf{x})$ through a sequence of $N_\beta$ intermediate annealing distributions
\begin{equation}
    \label{Eq:AIS_pi}
    p_i(\mathbf{x}) = p_0(\mathbf{x})^{1-\beta_i} p(\mathbf{x})^{\beta_i},
\end{equation}
with $0 = \beta_0 < \beta_1 < \dots < \beta_{N_\beta} = 1$, being $N_\beta$ the total number of annealing steps. This is done by direct propagation of $N_s$ Markov chains (or replicas) in parallel, each one contributing at the end with a weight $\omega_i$.

Using the sequence of configurations generated along the annealing path, expectation values of observables can be estimated as
\begin{equation}
    \label{Eq:AIS_average}
    \langle \mathcal{O} \rangle \approx \frac{\sum_{k=1}^{N_s} \tilde{\omega}_k \mathcal{O}_k}    {\sum_{k=1}^{N_s} \tilde{\omega}_k},
\end{equation}
where the unnormalized importance weights $\tilde{\omega}_k$ are given by
\begin{equation}
    \tilde{\omega}_k = \prod_{i=0}^{N_\beta-1} \frac{\tilde{p}_{i+1}(\mathbf{x}_{i}^{(k)})}{\tilde{p}_{i}(\mathbf{x}_{i}^{(k)})},
\end{equation}
with $\tilde{p}_i(\mathbf{v}, \mathbf{h})$ denoting the unnormalized intermediate distribution and the index $k$ labelling the replica. The partition function is then estimated from the relation
\begin{equation}
    \frac{Z}{Z_0} \approx \frac{1}{N_s} \sum_{k=1}^{N_s} \tilde{\omega}_k,
\end{equation}
where $Z_0$ is the partition function of $p_0({\bf x})$, which is assumed to be known.

\subsection{Improving AIS performance on RBMs}
\label{Subsec:AIS_improvement}

In general, the simplest choice is to start AIS from the uniform probability distribution, corresponding to an infinite temperature Boltzmann distribution. While convenient, this initialization can lead to inaccurate estimates of the partition function for systems presenting complex energy landscapes. In the particular case of an RBM, it is very easy to improve the initialization by considering an initial distribution of the form
\begin{equation}
    p_0(\mathbf{x}) \propto e^{\mathbf{x}^T \mathbf{B}/T},
\end{equation}
where $\mathbf{B}$ is a a vector of mean fields, chosen to minimize the Kullback--Leibler divergence between $p_0(\mathbf{x})$ and the target distribution $p(\mathbf{x})$. For binary $\{0,1\}$ units, this results in
\begin{equation}
   \mathbf{B} = -T \ln \left( \frac{1}{\langle \mathbf{x} \rangle}_{p({\bf x})} - 1 \right) \ ,
\end{equation}
while
\begin{equation}
    \mathbf{B} = -T \tanh^{-1} \left( \langle \mathbf{x}_{p({\bf x})} \rangle \right) \ .
\end{equation}
for binary $\{-1, 1\}$ ones. In both expressions, $\langle \mathbf{x} \rangle_{p({\bf x})}$ stands for the expected value of $\mathbf{x}$ under the target distribution $p(\mathbf{x})$. Notice that setting $\mathbf{B} = \mathbf{0}$ recovers the uniform distribution. In the following this will be referred to as AIS-\textit{B0}.

Since the exact evaluation of $\langle \mathbf{x} \rangle_{p({\bf x})}$ is as difficult as the computation of the partition function, a practical approximation is required. In this work, we employ the heuristic known as \textit{Signs from Random Hiddens} (AIS-\textit{Sgh}), originally introduced in~\cite{Article_original}. The visible and hidden mean values are estimated separately. With ${\bf x}$ the tuple of visible and hidden unit configurations $({\bf v},{\bf h})$, one starts by drawing a set of hidden states ${\bf h}^{(\alpha)}$ and getting from them the corresponding value of the visible units that locally minimize the RBM energy of Eq.~(\ref{eq:RBM_ene}), obtained by setting
\[
v_i^\alpha = 
\begin{cases}
1, & {\rm if\,\,\,}b_i + \mathbf{W}_i \mathbf{h}^\alpha > 0, \\
0 {\rm \,\,\,or\,\,\,}-1 , & \text{otherwise},
\end{cases}
\]
the last condition depending on whether units are $\{0,1\}$ or $\{-1,1\}$. $\mathbf{W_i}$ corresponds to the $i-{th}$ row of $\mathbf{W}$. The mean $\langle \mathbf{v} \rangle$ is then approximated by averaging over the resulting configurations $\{\mathbf{v}^\alpha\}$.

In order to obtain $\langle \mathbf{h} \rangle$, we follow the same procedure but generating a set of random visible units ${\bf v}^{(\alpha)}$ and obtaining their corresponding hidden values as
\[
h_i^\alpha = 
\begin{cases}
1, & {\rm if\,\,\,}c_i + (\mathbf{v}^\alpha)^T\mathbf{W}_i  > 0, \\
0 {\rm \,\,\,or\,\,\,}-1 , & \text{otherwise}.
\end{cases}
\]

The AIS-\textit{Sgh} heuristic was originally validated in~\cite{Article_original}. The present work extends its application to a broader class of systems and, crucially, demonstrates its utility as a quantitative tool for phase transition detection and finite-size scaling analysis, which was not addressed previously.

\section{Results}
\label{Sec:Results}
This section presents our main results, organized around two complementary goals. In Sec.~\ref{Subsec:Results_AISB0}, we benchmark how AIS-\textit{Sgh} detects and describes phase transitions on well-studied spin systems — the 2D and 3D Ising model~\cite{Ising}, the Edwards–Anderson spin-glass~\cite{Edwards_1975}, and the bipartite Sherrington-Kirkpatrick spin-glass model~\cite{BipartiteSK_model} — where reliable reference values exist in the literature. This serves as a quantitative validation of the method. In Sec.~\ref{Subsec:Results_AISSgh}, we move to RBMs with increasingly complex energy landscapes, including both artificial constructions and networks trained on real data. Together, these two parts build a progressive case for AIS-\textit{Sgh} as a robust and general tool for thermodynamic analysis across a wide class of spin-like models.

\subsection{Criticality on Ising and Spin-glass systems}
\label{Subsec:Results_AISB0}
These systems will serve as a benchmark to study how AIS identifies and predicts critical temperatures exponents. We then compare the values found with AIS to the ones reported in the literature.

All systems analyzed in this section use $\{-1,+1\}$ spin values and no external magnetic field. Under these conditions the energy is invariant under the $x_i \to -x_i$ transformation, and thus the average magnetization is $\langle m \rangle = 0$. Any non-zero bias $\mathbf{B}$ in the AIS initial distribution would artificially break this symmetry, making the uniform initialization AIS-\textit{B0} the natural and formally optimal choice. Nevertheless, we will deliberately use both AIS-\textit{B0} and AIS-\textit{Sgh} in the following tests, in order to showcase that AIS-\textit{Sgh} is able to estimate critical temperatures and exponents with the same level of precision than AIS-\textit{B0}, even in a regime where it offers no theoretical advantage.

Gibbs sampling is used in all evaluations, and statistical averages are taken over both visible and hidden units, as in this context they both play the role of magnetic spins, spanning each one half the number of spins of the ensemble considered. Statistical uncertainties on all estimated quantities were computed using the jackknife resampling method~\cite{Jackknife}.

\subsubsection{Ising systems}
\label{Subsec:Results_Ising}

We used AIS to study the magnetization and internal energy of the 2D and 3D Ising models, comparing the results with those obtained using MH and GS. We considered a $50 \times 50$ lattice for the 2D case and a $12 \times 12 \times 12$ lattice for the 3D case.

\begin{figure}
    \centering

    \includegraphics[width=0.8\linewidth]{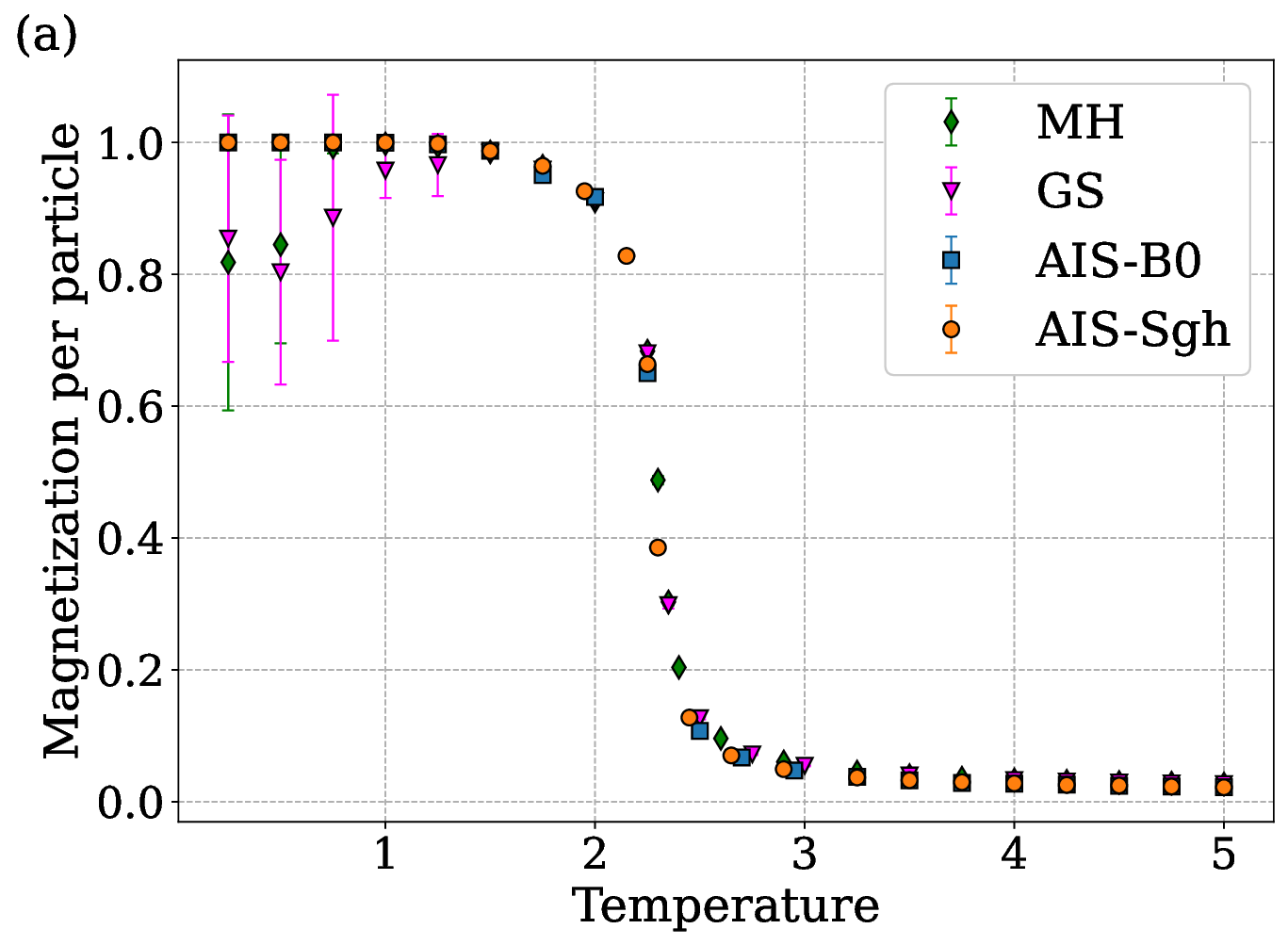}
    \includegraphics[width=0.8\linewidth]{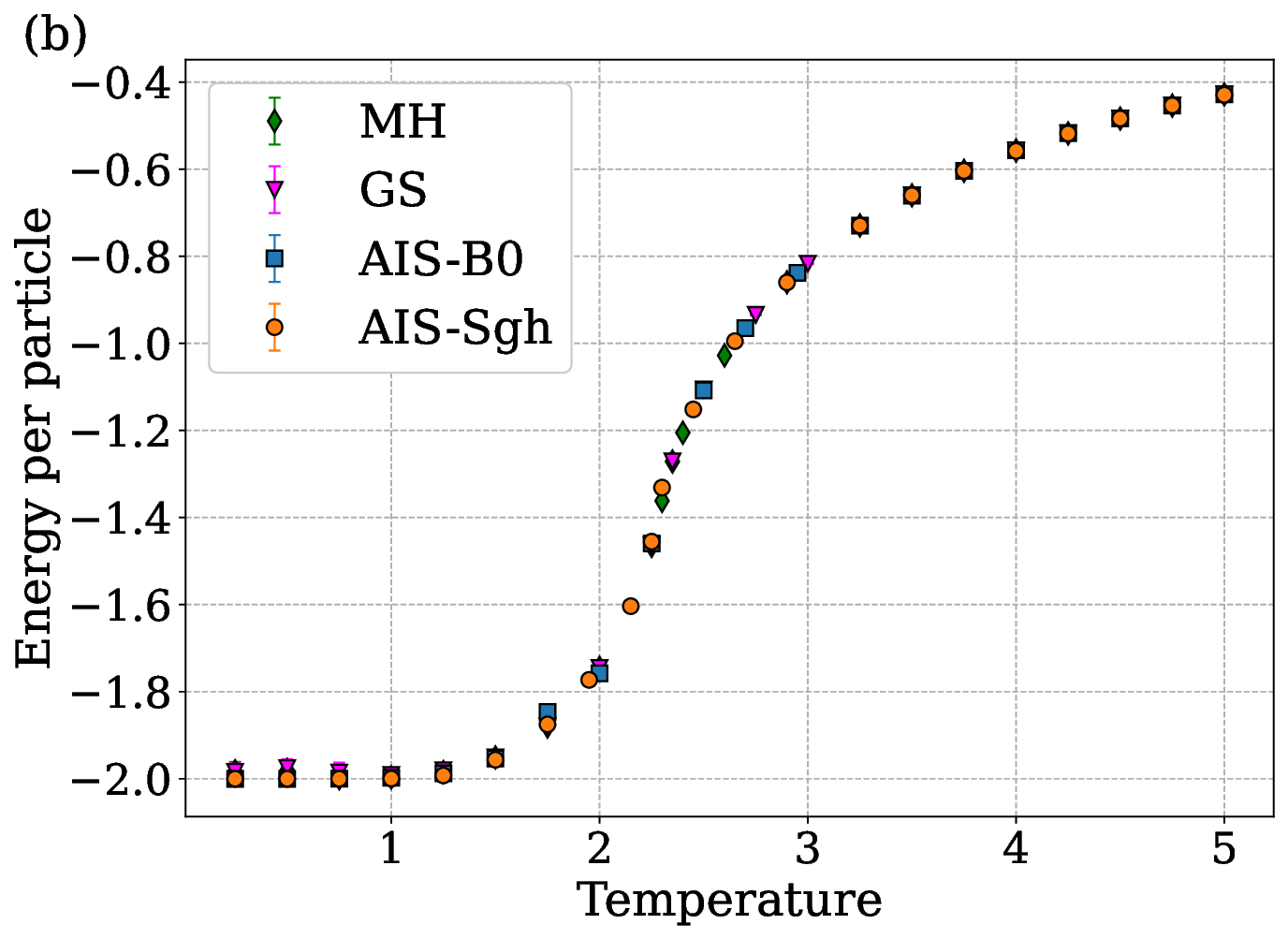}
    \includegraphics[width=0.8\linewidth]{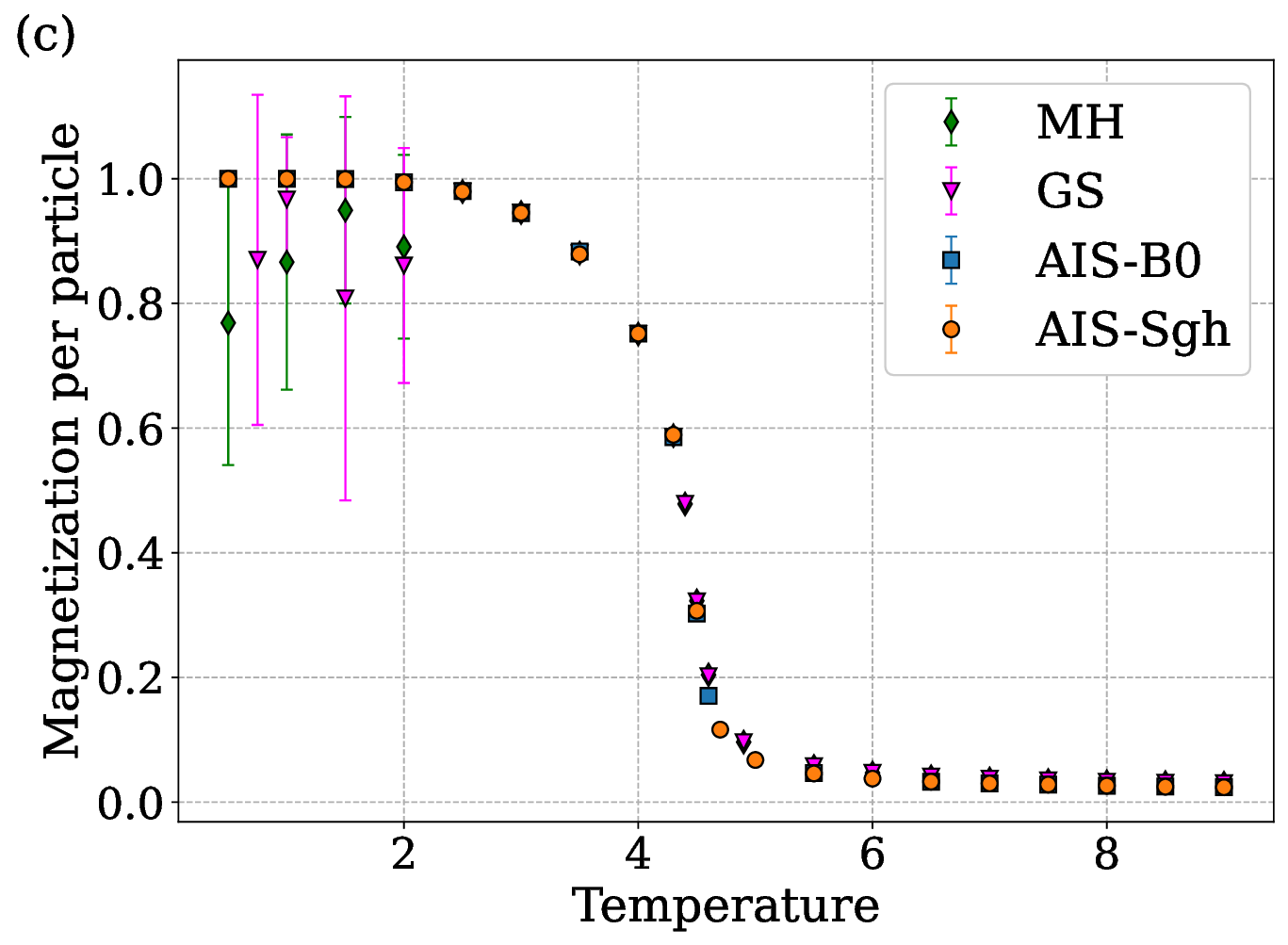}
    \includegraphics[width=0.8\linewidth]{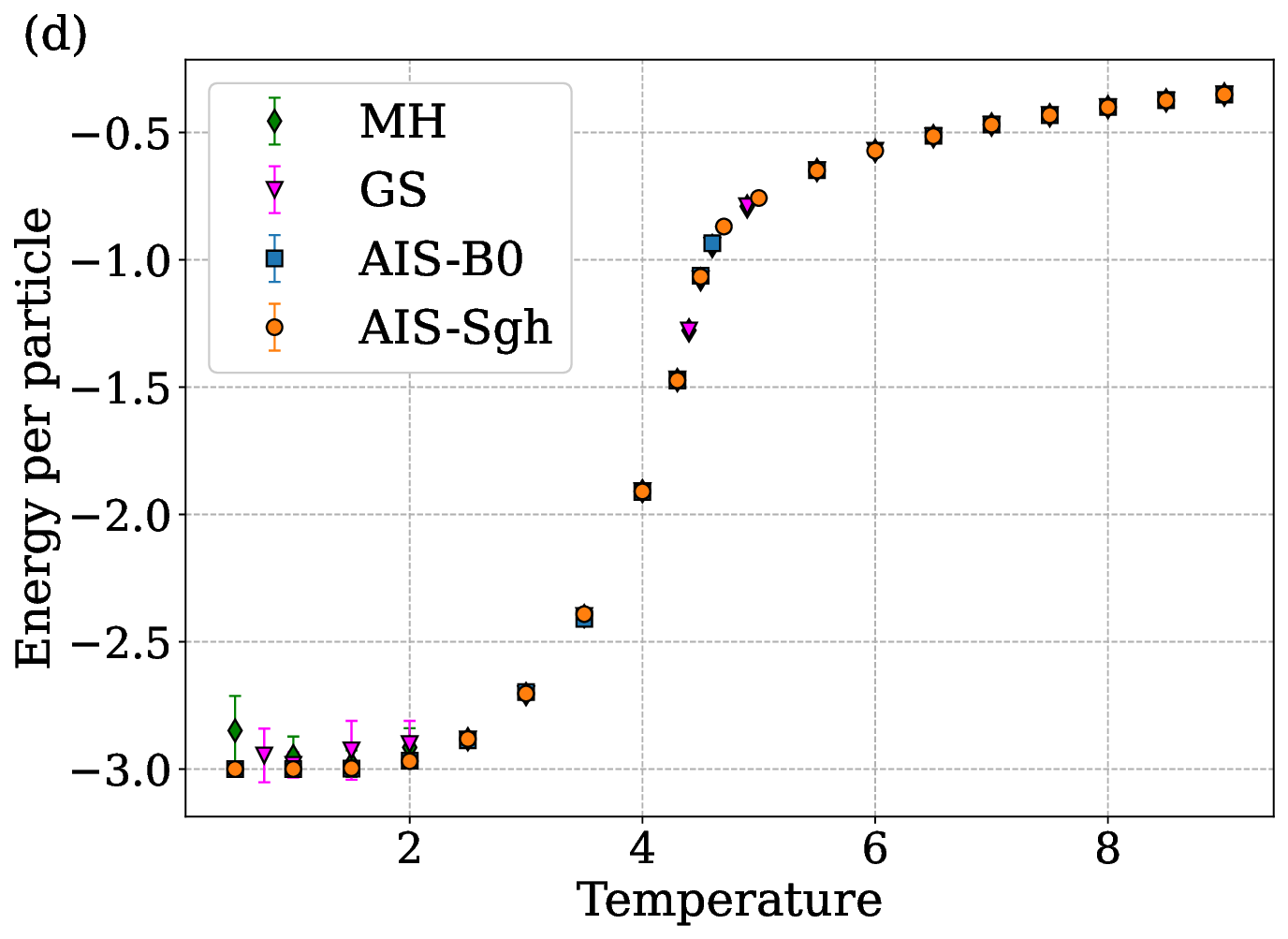}

    \caption{Magnetization and internal energy per particle of the 2D $50 \times 50$ (panels (a) and (b)) and 3D $12 \times 12 \times 12$ (panels (c) and (d)) Ising models at different temperatures. Results are shown for Metropolis--Hastings (MH), Gibbs Sampling (GS), and Annealed Importance Sampling using both initializations (AIS-\textit{B0} and AIS-\textit{Sgh}). For MH and GS, the system was thermalized for $4096$ Monte Carlo steps and $4096$ samples were collected per temperature. For AIS, $N_\beta = N_s = 4096$.}
    \label{fig:magne_ene_Ising2n3D}
\end{figure}

For MH and GS, the system was thermalized for $4096$ Monte Carlo steps, after which $4096$ samples were collected at each temperature to compute ensemble averages. For AIS, we used $N_\beta = 4096$ and $N_s = 4096$.

The results are shown in Fig.~\ref{fig:magne_ene_Ising2n3D}. As it can be seen, both AIS initializations provides more accurate estimates than both MH and GS in the low-temperature regime. With a finite number of Monte Carlo steps, the Markov chains may become trapped in local energy minima, leading to biased estimates of thermodynamic observables~\cite{AIS_Millora_SMC}. AIS, on the other hand, is able to escape those local minima and provide accurate representation of the magnitudes at this low-temperature regime.

\begin{figure}
    \centering
    \includegraphics[width=0.95\linewidth]{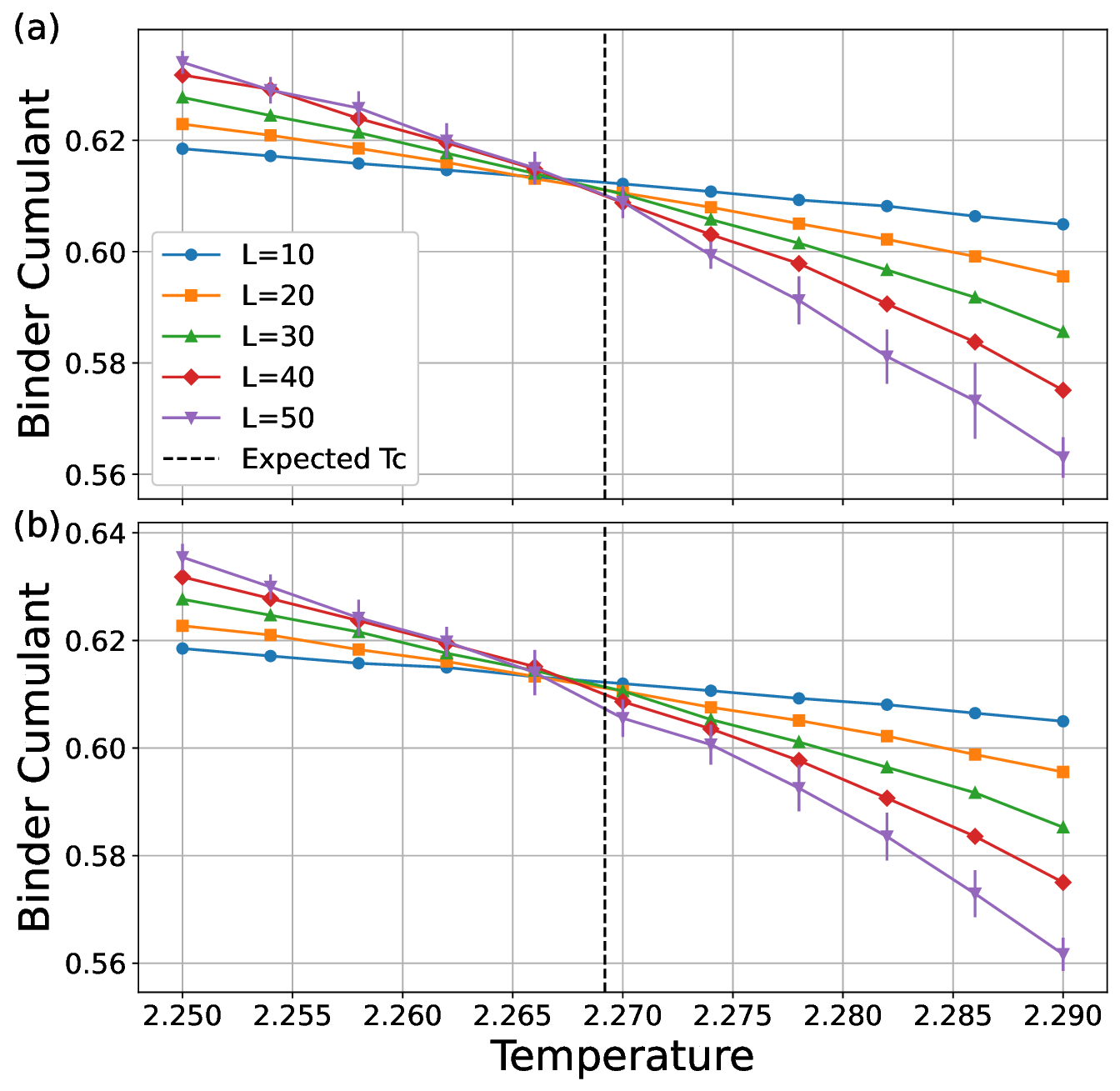}
    \caption{Binder cumulant as a function of temperature for the 2D Ising model for different lattice sizes $L$ using AIS-\textit{B0} (a) and AIS-\textit{Sgh} (b). The crossing point of the curves estimates the critical temperature $T_c$. Lines are guides to the eye.}
    \label{fig:Ising2D_Binder}
\end{figure}

\begin{figure}
    \centering
    \includegraphics[width=0.95\linewidth]{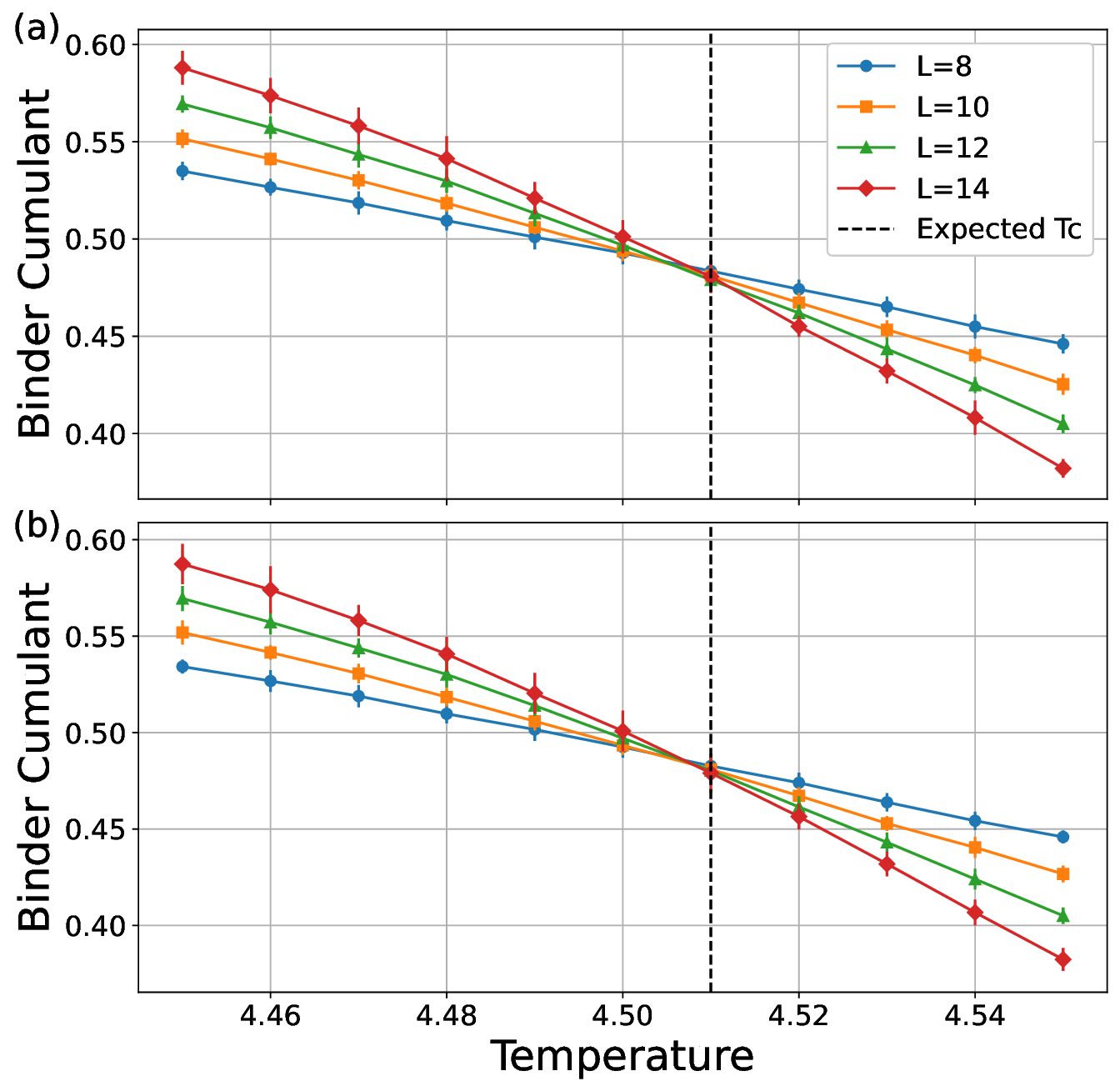}
    \caption{Binder cumulant as a function of temperature for the 3D Ising model for different lattice sizes $L$ using AIS-\textit{B0} (a) and AIS-\textit{Sgh} (b). The crossing point of the curves estimates the critical temperature $T_c$. Lines are guides to the eye.}

    \label{fig:Ising3D_Binder}
\end{figure}

Our next goal was to assess the performance of AIS in detecting and characterizing a phase transition. The Ising model undergoes a ferromagnetic phase transition in both two and three dimensions. For $d = 2$, Onsager derived an exact solution in the thermodynamic limit~\cite{Onsager}. For $d = 3$, high-precision numerical studies have obtained reliable values for both the critical temperature and the critical exponents~\cite{Ising_3D, Ising3D_2}.

As the standard procedure, we estimate the critical temperature using the Binder cumulant~\cite{BinderCumulant},
\begin{equation}
\label{Eq:binder}
    U_4 = 1 - \frac{\langle \Psi^4 \rangle}{3 \langle \Psi^2 \rangle^2},
\end{equation}
where $\Psi$ is to the order parameter of the system. In the Ising model, this order parameter corresponds to the magnetization per spin $m$ from Eq.~(\ref{Eq:magne}). For continuous phase transition, $U_4$ becomes independent of system size at $T = T_c$ in the thermodynamic limit, up to finite-size corrections. The critical temperature is then estimated from the crossing point of the $U_4$ curves for different lattice sizes~\cite{MC_SP1,FSS,FSS2}.

We estimated the critical exponents $\nu$, $\gamma$, and $\beta$, defined as $\xi \propto |T-T_c|^{-\nu}$, $\chi \propto |T - T_c|^{-\gamma}$ and
$\Psi \propto |T-T_c|^\beta$, being $\xi$ the correlation length and $\chi$ the susceptibility. Those exponents were extracted using standard finite-size scaling relations~\cite{MC_SP1}. The exponent $\nu$ is obtained from the slope of the Binder cumulant at the critical point, which scales as $dU_4/dT\big|_{T_c} \propto L^{1/\nu}$; plotting this slope as a function of $L$ on a log-log scale yields $\nu$ from the slope of the resulting line. Similarly, $\gamma$ is extracted from the finite-size scaling of the magnetic susceptibility at the critical point, $\chi(T_c) \propto L^{\gamma/\nu}$, and $\beta$ from the scaling of the magnetization, $m(T_c) \propto L^{-\beta/\nu}$. The magnetic susceptibility was obtained using
\begin{equation}
    \chi = N\frac{\langle m^2\rangle - \langle m \rangle^2}{T}~,
\end{equation}
where the averages were obtained using Eq.~(\ref{Eq:AIS_average}).

For the 2D case, we use square lattices of sizes $L \in \{10, 20, 30, 40, 50\}$ with $N = L^2$ spins, while the 3D case $L \in \{8, 10, 12, 14\}$ with $N = L^3$ spins. In all these simulations, we set $N_\beta = N_s = 4096$. The Binder cumulant curves are shown in Fig.~\ref{fig:Ising2D_Binder} for the 2D case and in Fig.~\ref{fig:Ising3D_Binder} for the 3D case, and the resulting critical parameters are summarized in Table~\ref{tab:Ising_2D} (2D) and in Table~\ref{tab:Ising_3D} (3D).

\begin{table}[ht]
\centering
\caption{Comparison of the results obtained with AIS-\textit{B0} and AIS-\textit{Sgh} of the different critical values of the Ising model in 2D with the exact results derived by Onsager~\cite{Onsager}. The errors were calculated using the jackknife method.}
\begin{ruledtabular}
\begin{tabular}{cccc}
         & AIS-\textit{B0}     & AIS-\textit{Sgh}    & Onsager   \\ \hline
$T_c$    & $2.267 \pm 0.001$   & $2.266 \pm 0.003$   & $2.269...$ \\
$\nu$    & $1.05 \pm 0.05$     & $1.04 \pm 0.02$     & $1$        \\
$\gamma$ & $1.83 \pm 0.10 $    & $1.82 \pm 0.07$     & $1.75$     \\
$\beta$  & $0.128 \pm 0.009 $  & $0.121 \pm 0.003 $  & $0.125$    \\ \hline
\end{tabular}
\end{ruledtabular}
\label{tab:Ising_2D}
\end{table}

\begin{table}[ht]
\centering
\caption{Comparison of the results obtained with AIS-\textit{B0} and AIS-\textit{Sgh} of the different critical values of the Ising model in 3D with the results found in the literature~\cite{Ising_3D,Ising3D_2}. The errors were calculated using the jackknife method.}
\begin{ruledtabular}
\begin{tabular}{cccc}
         & AIS-\textit{B0}    & AIS-\textit{Sgh}    & Literature    \\ \hline
$T_c$    & $4.508 \pm 0.005$  &  $4.507 \pm 0.017$  & $4.511...$    \\
$\nu$    & $0.62\pm 0.05$     &  $0.64\pm 0.04$     & $0.62997...$  \\
$\gamma$ & $1.28 \pm 0.09$    &  $1.29 \pm 0.07$    & $1.2370...$   \\
$\beta$  & $0.30 \pm 0.03 $   &  $0.300 \pm 0.017 $ & $0.3264...$   \\ \hline
\end{tabular}
\end{ruledtabular}
\label{tab:Ising_3D}
\end{table}

In all cases, the AIS estimates are consistent with the literature values to within the quoted uncertainties, confirming that AIS is capable of accurately resolving the critical behavior of the Ising model in both two and three dimensions. Notice also how both initializations are able to obtain a similar estimation of those magnitudes, showing how AIS-\textit{Sgh} is able to match the performance of AIS-\textit{B0} even in a disadvantageous situation.

\subsubsection{Spin glass Ising Lattice}
\label{Subsec:Results_SG}
Next we test AIS on a more challenging system, the Edwards--Anderson (EA) spin-glass model~\cite{Edwards_1975} on a simple cubic lattice with coordination number 6, where interactions between nearest neighbors are assigned a couplings $\pm J$ with probability $p = 0.5$. This system is known to exhibit a spin-glass to paramagnetic phase transition at $T_c \approx 1.1019$~\cite{EA3D_1, EA3D_2, EA3D_3}.

To characterize the spin-glass phase, Edwards and Anderson introduced an order parameter based on the overlap between two replicas of the same system,
\begin{equation}
    q^{(a,b)} = \frac{1}{N} \sum_i x_i^{(a)} x_i^{(b)},
\end{equation}
where $i$ labels the lattice sites, $a$ and $b$ denote different replicas, and $N$ is the total number of spins. When more than two replicas are available, the mean overlap $\langle q \rangle$ can be obtained by averaging over all distinct replica pairs, similarly to~\cite{qEA_replicas1,qEA_replicas2,qEA_replicas3}. Throughout this section, we will use this parameter as the order parameter.

Computing $q^{(a,b)}$ within AIS requires some care, since each AIS chain produces a single configuration. We simulated $N_s = 1024$ independent chains, each yielding one configuration, and paired all distinct chains to compute the overlap. Since each chain $a$ carries an importance weight $\omega_a$ (see 
Eq.~(\ref{Eq:AIS_average})), the thermally weighted estimator for $\langle q 
\rangle$ is
\begin{equation}
    \langle q \rangle =
    \frac{\displaystyle\sum_{a=1}^{N_s-1} \sum_{b=a+1}^{N_s}
    q^{(a,b)} \, \omega_a \omega_b}
    {\displaystyle\sum_{a=1}^{N_s-1} \sum_{b=a+1}^{N_s}
    \omega_a \omega_b}.
\end{equation}

The derivation of this expression can be found in the supplementary material.

We used this order parameter to compute the Binder cumulant $U_4$ (Eq.~(\ref{Eq:binder})) as a function of temperature for several lattice sizes. For each lattice size and temperature, we generated $N_{J_{ij}}$ independent disorder realizations of the couplings $J_{ij}$. For each realization, we computed $\langle q \rangle$, $\langle q^2 \rangle$ and $\langle q^4 \rangle$ using the weighted AIS samples and then averaged over disorder to obtain $U_4$. We used lattice sizes $L \in \{6, 8, 10, 12\}$, with $N_{J_{ij}} = 2000$ for all $L$. Notice that those are the most computationally expensive calculations done in this work.

\begin{figure}[h]
    \centering
    \includegraphics[width=0.95\linewidth]{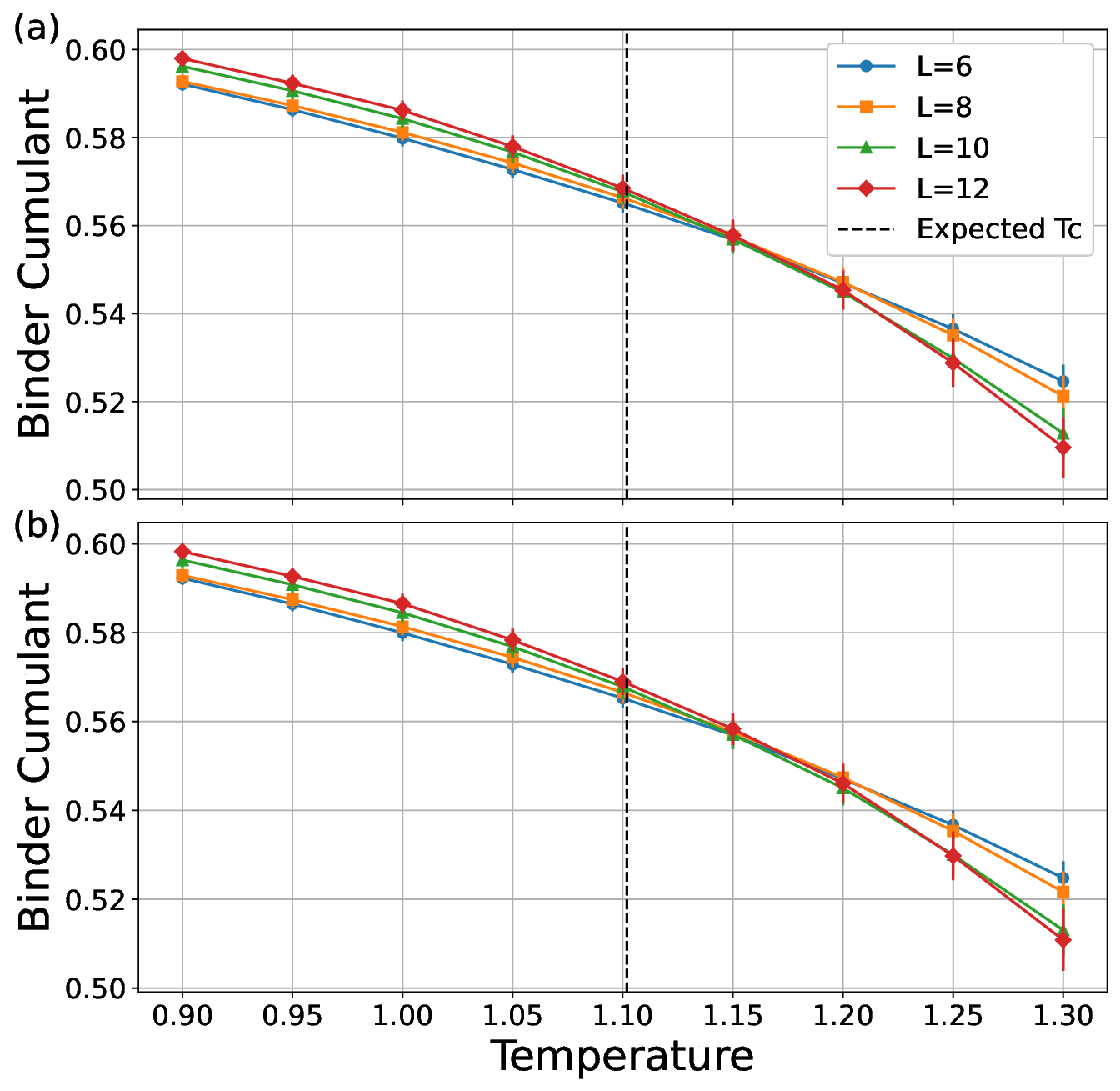}
    \caption{Binder cumulant as a function of temperature for the 3D $\pm J$ Edwards--Anderson model for different lattice sizes $L$. The crossing point of the curves estimates the critical temperature $T_c$. Panel (a) corresponds to the results using AIS-\textit{B0}, while panel (b) are the results for AIS-\textit{Sgh}.}
    \label{fig:EA3D_Binder}
\end{figure}

The resulting Binder cumulants are shown in Fig.~\ref{fig:EA3D_Binder}. As with the previous cases, the critical temperature was estimated from the crossing points of the curves, and the critical exponents from the finite size scaling relations. The obtained results are summarized in Table~\ref{tab:EA3D}.

\begin{table}[ht]
\centering
\caption{Results of the critical values of the EA 3D spin glass model using AIS-\textit{B0} and AIS-\textit{Sgh} compared with the ones found in the literature~\cite{EA3D_1,EA3D_2}.}
\begin{ruledtabular}
\begin{tabular}{cccc}
          & AIS-\textit{B0}     & AIS-\textit{Sgh}   & Literature  \\ \hline
$T_c$     & $1.104 \pm 0.011$   & $1.110 \pm 0.016$  & $1.1019...$ \\
$\nu$     & $2.45  \pm 0.08$    & $2.53  \pm 0.08$   & $2.562...$  \\
$\gamma$  & $5.9   \pm 0.3 $    & $6.1   \pm 0.3 $   & $6.13...$   \\
$\beta$   & $0.73  \pm 0.05 $   & $0.75 \pm 0.05$    & $0.782...$  \\ \hline
\end{tabular}
\end{ruledtabular}
\label{tab:EA3D}
\end{table}

As with the previous case, notice that both AIS-\textit{B0} and AIS-\textit{Sgh} provide similar results within their corresponding errorbars, showing again no disadvantage on using AIS-\textit{Sgh} even in a disadvantageous situation. Notice also how all critical parameters are consistent with the literature results, within their respective uncertainty range. This demonstrates that AIS is capable of resolving the spin-glass transition in disordered systems with local connectivity, where the competition between ferromagnetic and antiferromagnetic interactions makes conventional sampling significantly more demanding than in the Ising case.

\subsubsection{Bipartite Sherrington--Kirkpatrick model}
\label{Subsec:Results_SG_RBM}
Next we considered the Bipartite Sherrington--Kirkpatrick (BSK) model~\cite{BipartiteSK_model}. This spin-glass model has the same structure as an RBM, with two different layers of spins with random interactions between spins of different layers, and no inter-layer connections. Throughout this analysis, we fixed the ratio $N_v / N_h = 1$ and studied system sizes $N_v \in \{128, 256, 512\}$.

The coupling weights were drawn independently from $W_{ij} \sim \mathcal{N}\!\left(0, 1/\sqrt{N_v}\right)$, with no bias terms (external fields). The $1/\sqrt{N_v}$ scaling ensures that the energy per spin does not diverge in the thermodynamic limit, in analogy with the Sherrington--Kirkpatrick (SK) model of spin glasses~\cite{SK_model,BipartiteSK_model}. This model is known to yield a spin-glass phase transition at exactly $T_c = 1.0$, with critical exponents that differ substantially from those of the EA model due to the fully-connected nature of the interactions~\cite{RBM_thermo1}.

For each system size, we generated $N_{J_{ij}} = 100$ independent disorder realizations to obtain reliable disorder-averaged estimates of the studied observables. The spin-glass transition was characterized using the EA order parameter and the Binder cumulant, following the same methodology as in Sec.~\ref{Subsec:Results_SG}.

\begin{figure}
    \centering
    \includegraphics[width=0.95\linewidth]{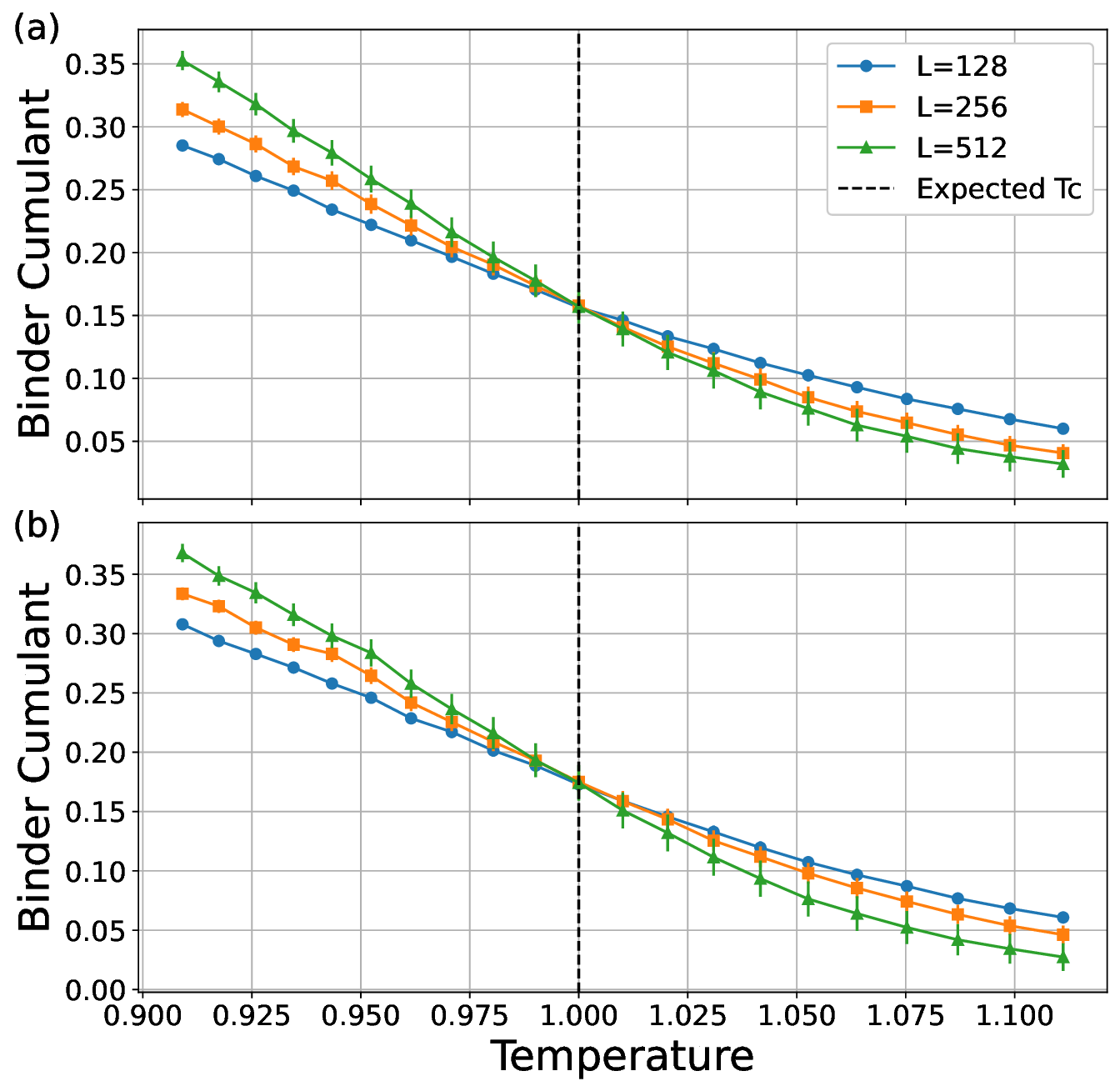}
    \caption{Binder cumulant as a function of temperature for the BSK model, for several system sizes $N_v$. Panel (a) corresponds to the results using AIS-\textit{B0}, while panel (b) are the results of AIS-\textit{Sgh}. The crossing of the curves signals the spin-glass phase transition.}
    \label{fig:RBM_SG}
\end{figure}

The Binder cumulant curves for all system sizes are shown in Fig.~\ref{fig:RBM_SG}. The crossing of the curves provides a clear signal of the phase transition, and the finite-size scaling analysis yields the critical values summarized in Table~\ref{tab:BSK}.

\begin{table}[ht]
\centering
\caption{Comparison between the results of the critical magnitudes obtained with AIS-\textit{B0} and AIS-\textit{Sgh} with their corresponding exact values~\cite{RBM_thermo1} for the BSK model.}
\begin{ruledtabular}
\begin{tabular}{cccc}
         & AIS-\textit{B0}     & AIS-\textit{Sgh}   & Exact     \\ \hline
$T_c$    & $1.001 \pm 0.003$  & $1.001  \pm 0.004$ & $1$        \\
$\nu$    & $3.09  \pm 0.10$    & $2.92  \pm 0.08$   & $3$        \\
$\gamma$ & $1.07  \pm 0.10 $   & $1.08  \pm 0.07 $  & $1$        \\
$\beta$  & $1.01  \pm 0.08 $   & $0.96  \pm 0.05 $  & $1$        \\ \hline
\end{tabular}
\end{ruledtabular}
\label{tab:BSK}
\end{table}

These values are in excellent agreement with the known critical values of this model, validating the accuracy of our AIS-based approach in the detection of spin-glass phase transitions in the BSK model.

These results establish AIS-\textit{Sgh} as a reliable tool for phase transition detection in systems with known critical behavior.

\subsection{Dependence on temperature}
\label{Subsec:Results_AISSgh}

We now turn to RBMs with complex energy landscapes. In these settings, no exact benchmark is available for large systems. The results in this section therefore serve a dual purpose: they stress-tests AIS-\textit{Sgh} in genuinely challenging regimes, and highlights the practical relevance of the initialization choice for reliable thermodynamic estimation in neural-network energy models.

In the following we compare the results obtained with AIS to those from PT. In all comparisons we use the same total number of samples $N_s = 1024$, with PT including $1024$ thermalization steps per replica, while in AIS we set $N_\beta = 1024$. This ensures that the comparison reflects genuine algorithmic differences in sampling quality rather than differences in computational budget, and provides a fair basis for evaluating the relative performance of the two methods.

For each problem we studied the different values of the magnetization $\langle m \rangle$, the EA parameter $\langle q \rangle$ and the partition function $Z$ using both methods and both AIS initializations, and compared them with their exact value in the cases when we had access to them. Since PT does not provide direct access to the partition function, only the AIS results are shown.

As in the previous section, GS is used to sample the intermediate probability distributions. On the other hand and for PT we decided to use MH since, after some test, we noticed that, although it's a slower sampling method, among the studied cases, it improved the global accuracy of the method.

As in the previous cases, the full visible and hidden system is employed to obtain the thermodynamical averages. The spins of those systems can take $\{0, 1\}$ values.

\subsubsection{Artificial RBMs}
\label{Subsec:Results_GWGM}

Following the previous spin glass systems, we have also tested a set of artificial neural networks with Gaussian-distributed weights, referred to as Gaussian Weights with Gaussian Moments (GWGM). The weight parameters were drawn from normal distributions $\mathcal{N}(\mu, \sigma)$, with mean $\mu$ and standard deviations $\sigma$ normally distributed, too, $\mu \sim {\mathcal N}(-10,10)$ and $\sigma \sim {\rm max}(0,{\mathcal N}(20,10))$. This procedure generates RBMs characterized by a single Gaussian mode with varying mean and variance.

We generated a total of 100 different RBMs. All networks consisted of $N_v = 20$ visible units and $N_h = 180$ hidden units. This relatively small system size allows for the exact computation of the partition function and magnetization, making GWGM systems particularly well suited for benchmarking sampling algorithms.

While Restricted Boltzmann Machines are known to exhibit a phase transitions in specific parameter regimes~\cite{RBM_thermo1, RBM_thermo2, RBM_train_PT}, here we focus on the thermodynamic properties of networks with fixed weights, rather than on phase transitions occurring during training. For these systems, we studied the partition function $Z$, the magnetization $\langle m \rangle$, and the EA parameter $\langle q \rangle$ as a function of the temperature. Exact analytical expressions for $Z$ and $\langle m \rangle$ can be easily derived and evaluated numerically, while the EA parameter can only be computed by {\em brute force}.

\begin{figure}
    \centering
    \includegraphics[width=\linewidth]{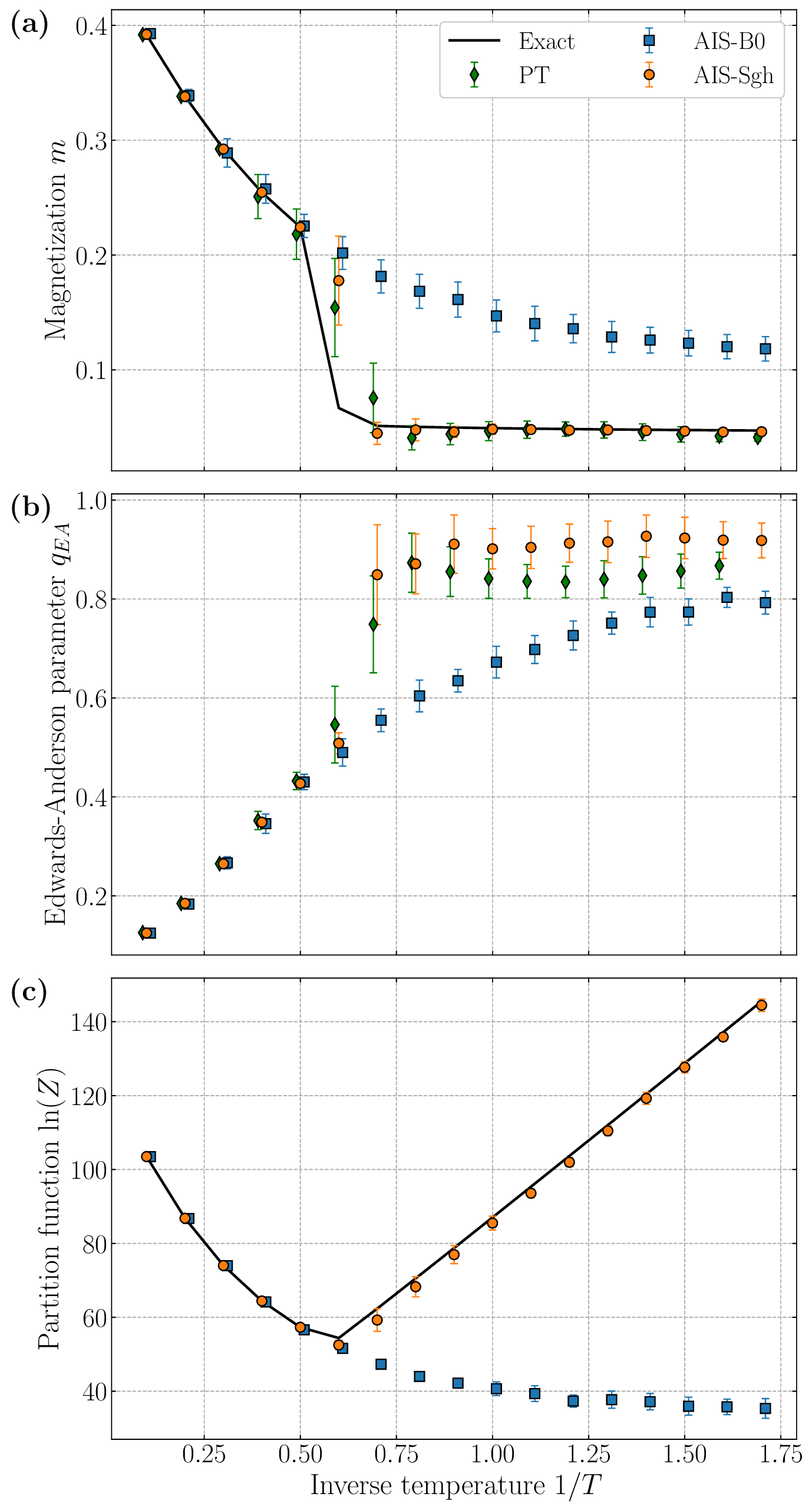}
    
    \caption{Comparison of AIS-\textit{B0}, AIS-\textit{Sgh}, and PT for a representative GWGM system. The panels show the magnetization $\langle m \rangle$ (a), the Edwards--Anderson parameter $\langle q \rangle$ (b), and the log-partition function $\ln Z$ (c) as functions of inverse temperature $1/T$. PT is excluded from panel (c) as it does not provide direct access to the partition function.}
    \label{fig:magne_Lokos87}
\end{figure}

Figure~\ref{fig:magne_Lokos87} shows representative results for a single GWGM instance, comparing the different mentioned methods. For PT, we used 17 evenly spaced temperatures spanning the interval $1/T \in [0.1,\, 1.7]$ where a change in behavior, associated to a phase transition, is clearly seen.

Examining the magnetization (panel (a)), both AIS-\textit{Sgh} and PT successfully detect the phase transition located at $1/T \approx 0.6$, although a small bias towards higher temperatures can be seen. This indicates that they are able to  correctly sample both the paramagnetic and the ferromagnetic phases, in agreement with the exact result. In contrast, AIS-\textit{B0} accurately reproduces the paramagnetic phase but fails to capture the transition to the ferromagnetic phase. This behavior is not surprising since AIS-\textit{B0} starts from an infinite-temperature, paramagnetic states, that is extended in metastable form down to quite low temperatures. In contrast, AIS-\textit{Sgh} places the system near high-probability configurations from the outset, enabling AIS to correctly sample both the paramagnetic and ferromagnetic phases and to identify the phase transition.

The EA parameter (panel (b)) exhibits a behavior that is very much in line with the magnetization. Only AIS-\textit{Sgh} and PT captures the phase transition, yielding a transition temperature in agreement with that obtained from $\langle m \rangle$, while AIS-\textit{B0} remain confined to the paramagnetic phase.

A qualitatively identical picture emerges for the partition function (panel (c)). 
AIS-\textit{B0} fails to detect the ferromagnetic transition, whereas 
AIS-\textit{Sgh} accurately tracks the exact behavior across both phases, 
demonstrating the importance of a well-chosen initial distribution for reliable 
thermodynamic estimation in multi-phase systems.

Finally, we have also tested the performance of both AIS initializations and PT across the full set of GWGM models over a wide range of temperatures. More specifically, we have computed the magnetization using AIS-\textit{B0}, AIS-\textit{Sgh} and PT for all 100 GWGM weights at 100 evenly spaced temperatures in the interval $T \in [0.05,\, 5.00]$, yielding a total of $10{,}000$ evaluation points. The accuracy of each prediction is quantified by the relative error

\begin{equation}
    \varepsilon = \left| \frac{m_{\text{exact}} - m_{\text{AIS}}}{m_{\text{exact}}} \right|,
    \label{Eq:Relative_error}
\end{equation}
and is assumed to yield a {\em successful} estimation whenever $\varepsilon \leq 0.05$.  

\begin{figure}
    \centering
    \includegraphics[width=\linewidth]{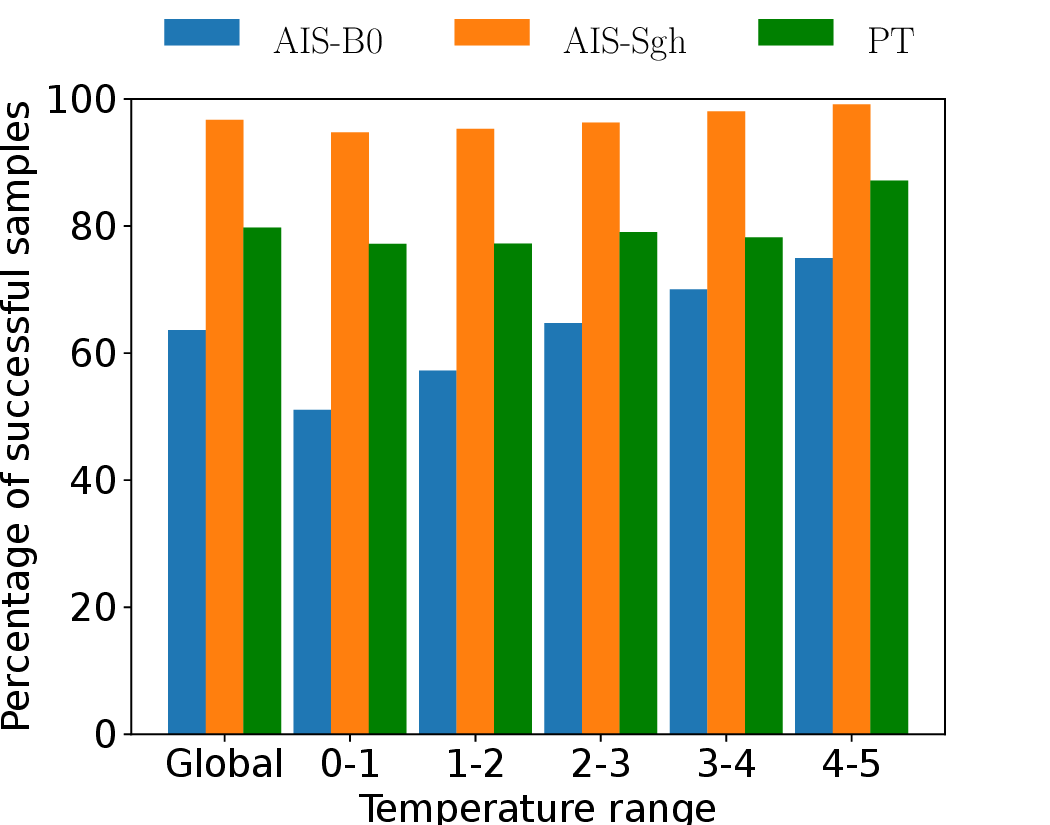}
    \caption{Success rate of the estimation of the magnetization $m$ for AIS-\textit{B0} (blue), AIS-\textit{Sgh} (orange) and PT (green), defined as the fraction of evaluations with relative error $\varepsilon \leq 0.05$ with respect to the exact value. Results are shown for 100 GWGM weight realizations at 100 temperatures in $T \in [0.05,\, 5.00]$, grouped by temperature interval; the leftmost bar aggregates all temperatures.}
    \label{fig:GWGM_PercentageSuccess}
\end{figure}

The results are summarized in Fig.~\ref{fig:GWGM_PercentageSuccess}, where each bar represents the success rate for a given initialization and temperature range. AIS-\textit{Sgh} consistently outperforms both AIS-\textit{B0} and PT across all temperature intervals with the established computational cost. As expected, both initializations improve at higher temperatures, where the system resides in the paramagnetic phase and sampling is less challenging. However, the performance gap between the two initializations is most pronounced at low temperatures, where AIS-\textit{B0} struggles to escape the paramagnetic basin while AIS-\textit{Sgh} retains a substantially higher success rate. These results confirm the robustness of the AIS-\textit{Sgh} initialization for reliable thermodynamic estimation across the full phase diagram of GWGM systems.

\subsubsection{Artificial RBMs with column restrictions}
\label{Subsec:Results_Rows}

\begin{figure}
    \centering
    \includegraphics[width=\linewidth]{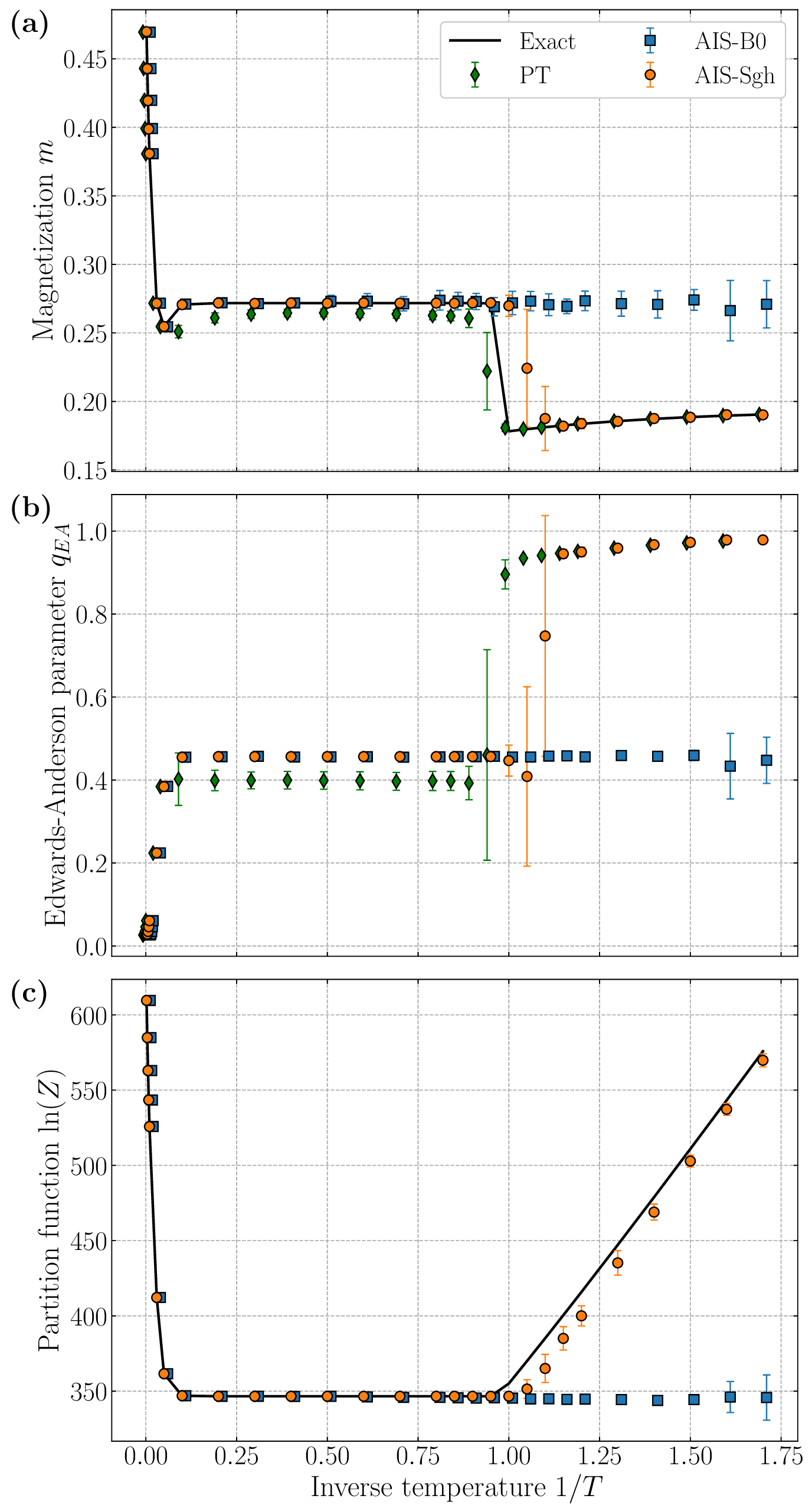}
    \caption{Comparison of AIS-\textit{B0}, AIS-\textit{Sgh}, and PT for a column-structured RBM with $k = 4$ cuts, $N_h = 420$ hidden units, and $N_v = 500$ visible units. Panels show the magnetization $\langle m \rangle$ (a), the Edwards-Anderson parameter $\langle q \rangle$ (b), and the log-partition function $\ln Z$ (c) as functions of inverse temperature $1/T$.}
    \label{fig:magne_Rows}
\end{figure}

We have also investigated a second class of artificial RBMs in which the interaction matrix exhibits a column-wise block structure. The weight matrix is organized into $k$ groups (or \emph{cuts}) of identical columns. In this sense, a weights matrix can schematically represented as

\[
W = 
\left(
\begin{array}{ccc|ccc|c|ccc}
a_1^{(1)}      & \dots  & a_1^{(1)}      & a_1^{(2)}      & \dots  & 
a_1^{(2)}      & \dots  & a_1^{(k)}      & \dots  & a_1^{(k)}      \\
a_2^{(1)}      & \dots  & a_2^{(1)}      & a_2^{(2)}      & \dots  & 
a_2^{(2)}      & \dots  & a_2^{(k)}      & \dots  & a_2^{(k)}      \\
\vdots         & \ddots & \vdots         & \vdots         & \ddots & 
\vdots         & \vdots & \vdots         & \ddots & \vdots         \\
a_{N_v}^{(1)} & \dots  & a_{N_v}^{(1)} & a_{N_v}^{(2)} & \dots  & 
a_{N_v}^{(2)} & \dots  & a_{N_v}^{(k)} & \dots  & a_{N_v}^{(k)}
\end{array}
\right).
\]

Due to the symmetry, the partition function and magnetization can be computed exactly and efficiently even for large systems, as long as $k$ is small. Denoting by $M_s$ the number of identical columns in the $s$-th cut, the partition function reduces to

\begin{multline}
    Z = \sum_{m_1=0}^{M_1} \sum_{m_2=0}^{M_2} \cdots \sum_{m_k=0}^{M_k}
    \binom{M_1}{m_1}\binom{M_2}{m_2} \cdots \binom{M_k}{m_k} \\
    \times \prod_{i=1}^{N_v}\left( 1 + e^{\,\sum_{s=1}^{k} m_s a_i^{(s)}} \right),
\end{multline}

while the magnetization is given by

\begin{multline}
    \langle m \rangle = \frac{1}{Z N}
    \sum_{m_1=0}^{M_1} \sum_{m_2=0}^{M_2} \cdots \sum_{m_k=0}^{M_k}
    \binom{M_1}{m_1}\binom{M_2}{m_2} \cdots \binom{M_k}{m_k} \\
    \times \left[
    \sum_{i=1}^{N_v} \sigma\!\left( \sum_{s=1}^{k} m_s a_i^{(s)} \right)
    \prod_{i=1}^{N_v}\!\left( 1 + e^{\,\sum_{s=1}^{k} m_s a_i^{(s)}} \right) \right. \\
    \left.
    + \sum_{s=1}^{k}\left(M_s - m_s\right)
    \prod_{i=1}^{N_v}\!\left( 1 + e^{\,\sum_{s'=1}^{k} m_{s'} a_i^{(s')} + a_i^{(s)}} \right)
    \right].
\end{multline}
These two expressions are derived in the supplementary material.

As with the GWGM case, we have generated 100 different weights sets with Gaussian random numbers, with $\mu$ and $\sigma$ also drawn from a random Gaussian distribution $\mu \sim {\mathcal N}(0,0.1)$ and $\sigma \sim {\rm max}(0,{\mathcal N}(0,0.1))$. All weights had $N_v = 500$ visible units and $N_h = 420$ hidden units arranged into $k = 4$ cuts of sizes $M_1 = 70$, $M_2 = 90$, $M_3 = 110$, and $M_4 = 150$, performing the same thermodynamic analysis as in the previous section. The results for a single, representative weight are shown in Fig.~\ref{fig:magne_Rows}.

Unlike the GWGM systems studied in the previous section, this model exhibits two distinct phase transitions: one corresponding to the paramagnetic-ferromagnetic transition of the visible units, and a second one associated with the analogous transition in the hidden layer. The overall picture differs qualitatively from the GWGM case in an important way: here, all three methods - AIS-\textit{B0}, AIS-\textit{Sgh}, and PT - successfully detect the higher-temperature transition and correctly sample both the paramagnetic phase and the intermediate phase. This is consistent with the smoother energy landscape at higher temperatures, which reduces the sampling difficulty for all algorithms.

The methods diverge, however, in their ability to resolve the lower-temperature
transition. AIS-\textit{B0} fails to detect it and does not correctly sample the low-temperature ferromagnetic phase. In contrast, both AIS-\textit{Sgh} and PT successfully identify the second transition and accurately sample all three phases across the full temperature range.

\begin{figure}
    \centering
    \includegraphics[width=\linewidth]{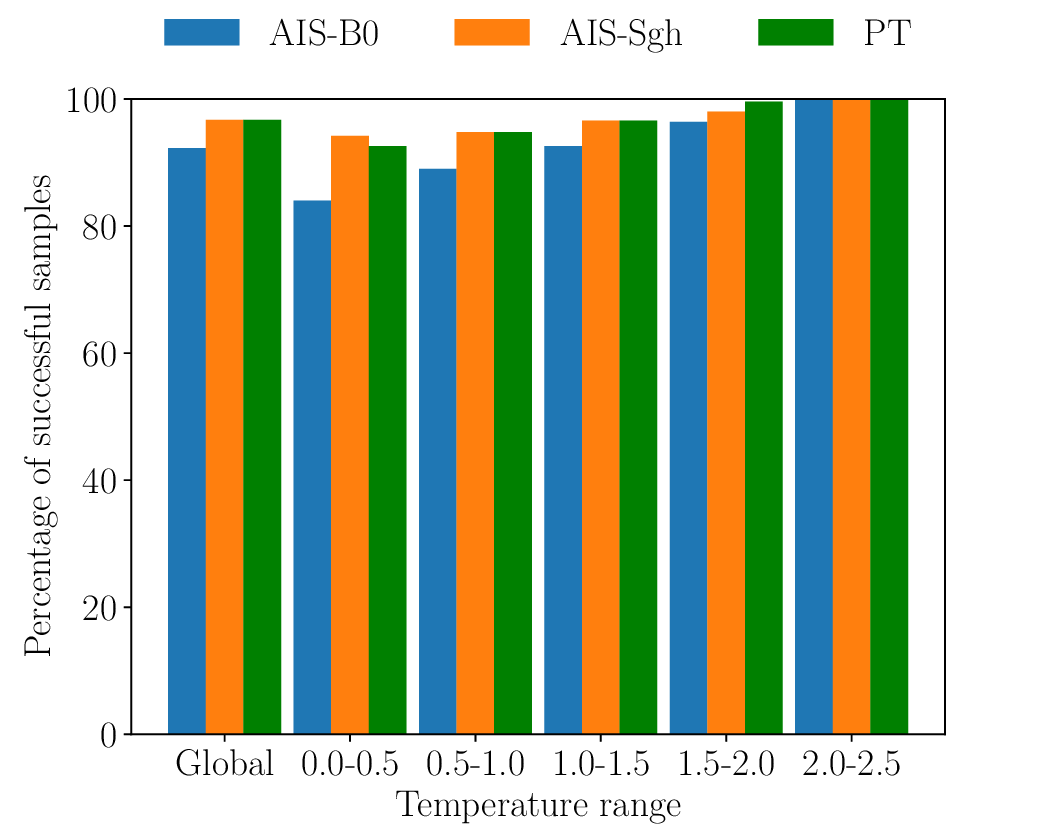}
    \caption{Percentage of success of the estimation of the magnetization for AIS-\textit{B0}, AIS-\textit{Sgh} and PT. A successful point corresponds to an evaluation with a relative error lower than 0.05 using Eq.~(\ref{Eq:Relative_error}). Results are shown for 100 column-restricted weights realizations at 50 different, evenly spaced temperatures $T\in[0.05, 2.5]$. The figure follows the same format as Fig.~\ref{fig:GWGM_PercentageSuccess}.}
    \label{fig:Rows_PercentageSuccess}
\end{figure}

As with the previous case with GWGM, we have also tested the performance of all methods and initializations across the full set of weights. In this case, we have tested the performance of AIS-\textit{B0}, AIS-\textit{Sgh} and PT across all the 100 different weights at 50 different temperatures evenly spaced between $T \in [0.05, 2.50]$, yielding a total of 5{,}000 evaluation points. The accuracy was also considered when the error using Eq.~(\ref{Eq:Relative_error}) is lower than $0.05$.

The results are summarized in Fig.~\ref{fig:Rows_PercentageSuccess}, which follows the same format as Fig.~\ref{fig:GWGM_PercentageSuccess}. AIS-\textit{Sgh} and PT obtain similar success rates, both globally and in each temperature group, whereas AIS-\textit{B0} is outperformed. The gap is largest at low temperatures, and shrinks progressively as $T$ increases, with all three methods reaching essentially $100\%$ at high temperatures.

\subsubsection{Trained RBMs}
\label{Subsec:Results_MNIST}

\begin{figure}
    \centering
    \includegraphics[width=\linewidth]{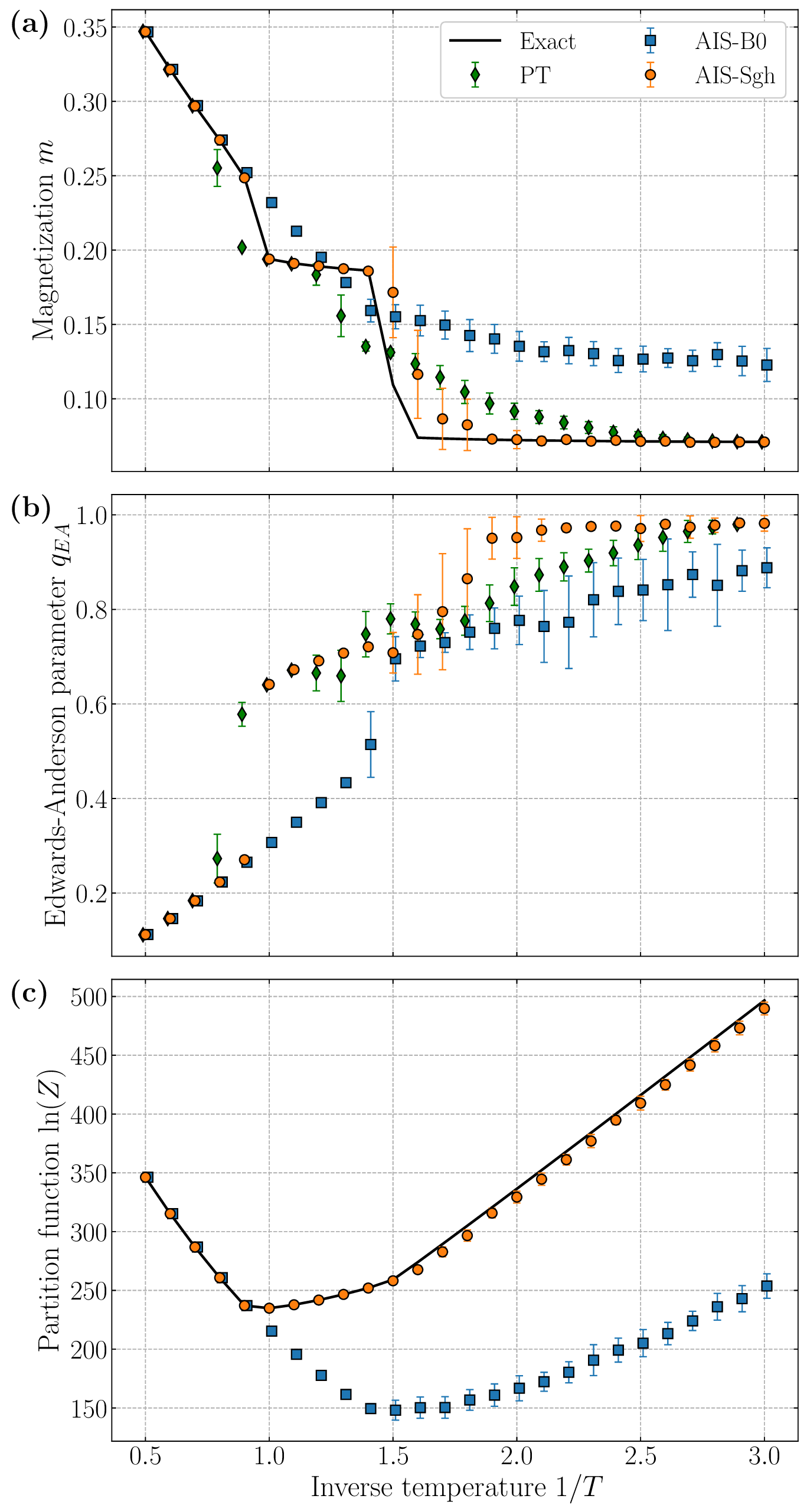}
    \caption{Comparison of AIS-\textit{B0}, AIS-\textit{Sgh}, and PT for an RBM trained on the MNIST dataset with $N_h = 20$ hidden units. Panels show the magnetization $\langle m \rangle$ (a), the Edwards--Anderson parameter $\langle q \rangle$ (b), and the log-partition function $\ln Z$ (c) as functions of inverse temperature $1/T$.}
    \label{fig:magne_MNIST20h}
\end{figure}

We finally consider two networks trained on the MNIST dataset~\cite{kaggleMNISTDataset,MNISTDataset}, one with $N_h = 20$ and the other with $N_h = 500$ hidden units, respectively, both containing $N_v=784$ visible nodes. Then, for the thermodynamic analysis, we used the learned weights at the end of training for the $N_h = 20$ network, and the weights at epoch 20 for the $N_h = 500$ network. We begin with the analysis of the smaller network, $N_h = 20$, for which the exact partition function and magnetization are accessible. Results for this system, computed using AIS-\textit{B0}, AIS-\textit{Sgh} and PT, are shown in Fig.~\ref{fig:magne_MNIST20h}

As it can be seen, all three methods track the exact magnetization curve accurately at high temperatures. Coming from this limit and lowering $T$, a phase transition is found at $1/T^{(1)}_{c} \approx 1$ as detected from a clear change in the behavior of $\ln{Z}$ around that point. This transition is neatly detected by both AIS-\textit{Sgh} and PT in the partition function, while the magnetization and the EA parameters show larger errors. On the other hand, AIS-\textit{B0} completely misses the transitions and extends in metastable form the wrong phase to lower temperatures. Moreover, AIS-\textit{Sgh} does a better job than PT (and AIS-\textit{B0}) at predicting the magnetization, although PT (with the computational cost considered) follows fairly well the exact curve.

The second transition, appearing at a lower temperature (around $1/T^{(2)}_c \approx 1.5$), proves to be more challenging for all methods. In this context AIS-\textit{B0} performs the worst, finding a transition but coming from the wrong phase, as seen in panel (c). On the other hand, PT detects the transition but systematically overestimates the magnetization at low temperatures, converging to a value noticeably above the exact result up to much lower temperatures. AIS-\textit{Sgh} is the only method that detects both transitions and tracks the correct low-temperature behavior: it exhibits the characteristic fluctuations and increased uncertainties expected near a critical point, but converges to a value consistent with the exact result. The larger error bars in this region are a natural consequence of the proximity to the critical point, where thermalization is intrinsically slower. In this way, AIS-\textit{Sgh} is able to detect phase transitions and produce reliable thermodynamic estimates even with a modest computational cost.

As with both previous cases, we now evaluate the accuracy of all three sampling methods across multiple weight configurations. Here, however, the set of weights considered are snapshots taken every ten epochs along training, thus capturing the full learning evolution. As learning is always done at a temperature $T=1$, we introduce an explicit temperature dependence by dividing the corresponding weights by a different $T$. For each snapshot, the magnetization at temperatures spanning the range $T \in [0.05,\, 2.50]$ is recorded, resulting in a total of $5{,}000$ evaluation points. The AIS estimates are then compared with the exact values using the relative error defined in Eq.~(\ref{Eq:Relative_error}).

\begin{figure}
    \centering
    \includegraphics[width=\linewidth]{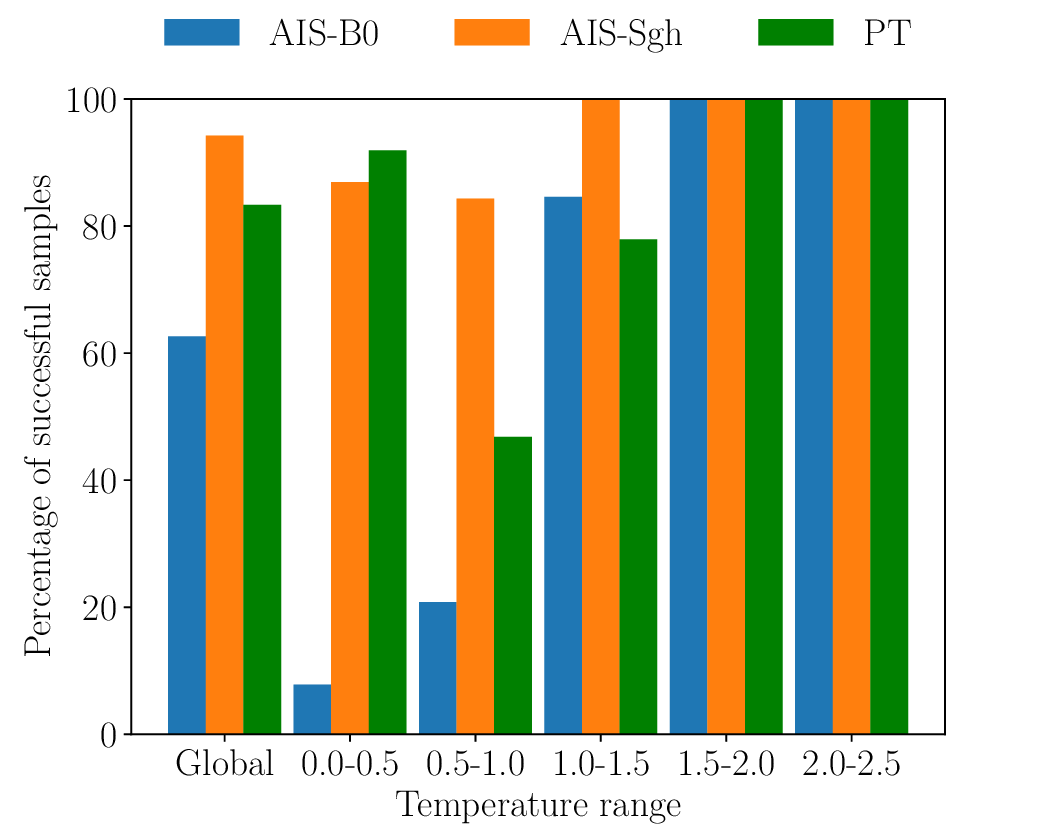}
    \caption{Success rate of the magnetization estimation of AIS-\textit{B0} (blue), AIS-\textit{Sgh} (orange) and PT (green), defined as the fraction of evaluation points with relative error $\varepsilon \leq 0.05$ using Eq.~(\ref{Eq:Relative_error}). Weights are snapshots taken every ten epochs during CD$_1$ training of an RBM on the MNIST dataset with $N_h = 20$ hidden units over $1{,}000$ epochs. Results are grouped by temperature interval; the leftmost bar aggregates all temperatures.}

    \label{fig:MNIST20h_PercentageSuccess}
\end{figure}

The results are summarized in Fig.~\ref{fig:MNIST20h_PercentageSuccess}, which follows the same format as in Fig.~\ref{fig:GWGM_PercentageSuccess}. As in the GWGM case, AIS-\textit{Sgh} outperforms AIS-\textit{B0} and PT at the global scheme. With the computational cost considered, though, PT is able to outperform AIS-\textit{Sgh} in the very low temperature regime $T \in [0, 0.5]$. At higher temperatures, where the system is fully paramagnetic, all sampling methods work equally well. These results indicate that AIS-\textit{Sgh} is the best overall performer in this problem also, with the fixed computational cost.

\begin{figure}
    \centering
    \includegraphics[width=\linewidth]{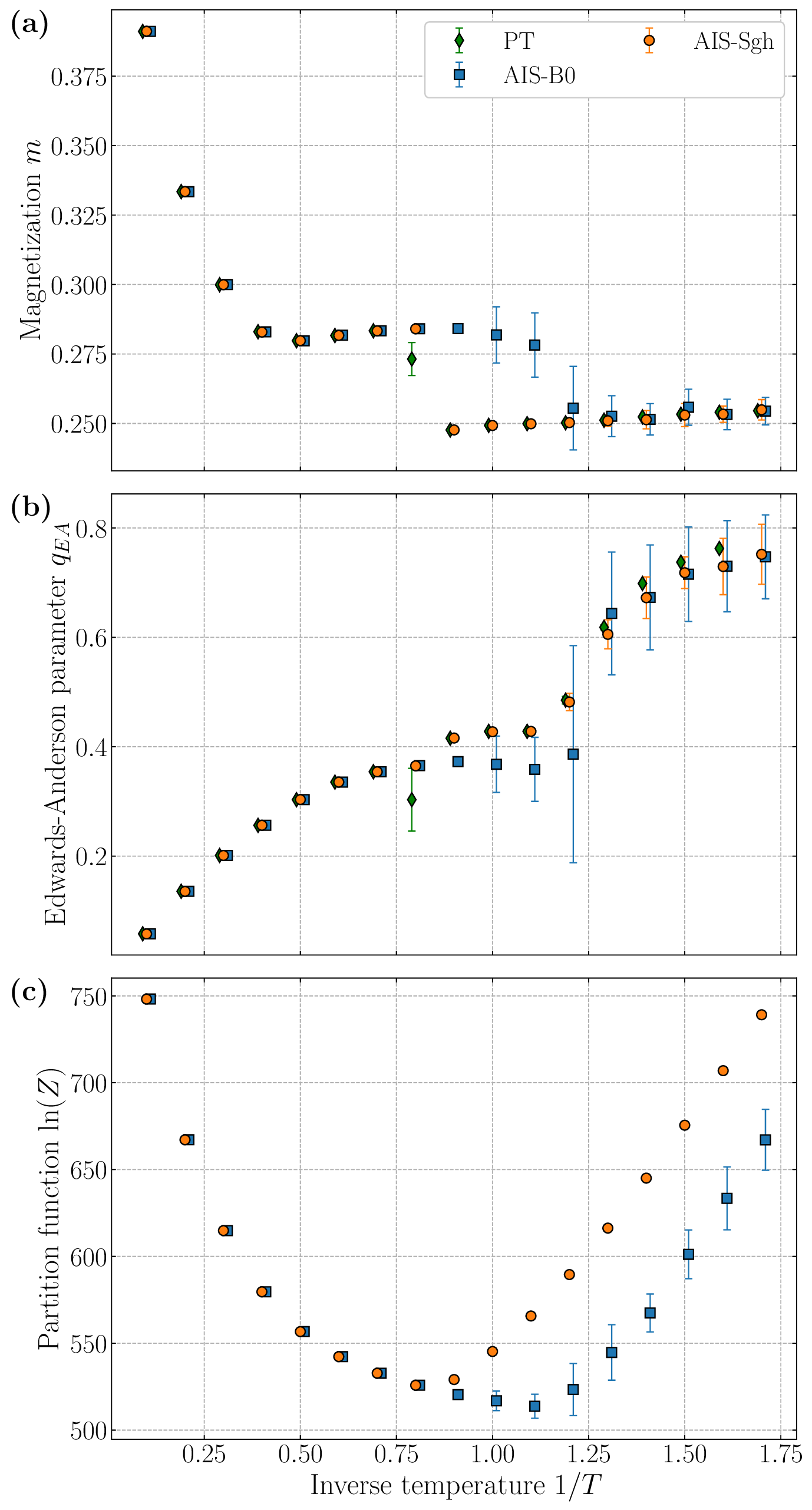}
    \caption{Comparison of AIS-\textit{B0}, AIS-\textit{Sgh}, and PT for an RBM trained on the MNIST dataset with $N_h = 500$ hidden units at epoch 20. Panels show the magnetization $\langle m \rangle$ (a), the Edwards--Anderson parameter $\langle q \rangle$ (b), and the log-partition function $\ln Z$ (c) as functions of inverse temperature $1/T$.}
    \label{fig:magne_MNIST500h}
\end{figure}

We now turn to the larger network trained on MNIST with $N_h = 500$ hidden units, which is the most difficult training problem considered in this work, and a standard benchmark in RBM learning. Due to the system size, the exact computation of any thermodynamic quantity is no longer feasible, and the comparison between methods must rely on their mutual consistency rather than on a ground-truth baseline. As before, we evaluate $\langle m \rangle$, $\langle q \rangle$, and $\ln Z$ as a function of the temperature using all sampling algorithms and initializations. We focus on a set of weights taken at an early stage of training (epoch 20): at later epochs, the model is more strongly constrained by the data and the differences between AIS-\textit{B0} and AIS-\textit{Sgh} become less pronounced, making the early-training regime the most informative one. The obtained results are shown in Fig.~\ref{fig:magne_MNIST500h}.

Regarding the magnetization, both AIS-\textit{Sgh} and PT detect a phase transition around $1/T \approx 0.8$, while AIS-\textit{B0} suggests a transition at a substantially lower temperature, ocurring around $1/T \approx 1.2$. This is consistent with the pattern observed in the $N_h = 20$ case, where AIS-\textit{B0} systematically underestimated the transition temperature by failing to escape the paramagnetic regime. These differences, however visible, are less pronounced in the EA parameter displayed on panel (b). The agreement between AIS-\textit{Sgh} and PT on the transition temperature lends additional credibility to the AIS-\textit{Sgh} estimate in this setting where no exact reference is available.

Regarding the partition function on panel (c), AIS-\textit{B0} systematically underestimates $\ln Z$ in the low-temperature regime, while AIS-\textit{Sgh} produces consistently higher values. In all previous cases where the exact partition function was accessible, the higher estimate proved to be the correct one. This is also what is expected on general grounds, as AIS tends to underestimate $\ln Z$ (see Ref.~\cite{Burda2015}). The larger estimate, which here corresponds to AIS-\textit{Sgh}, is therefore likely to be closer to the true value. Together, these two observations provide indirect evidence that AIS-\textit{Sgh} remains accurate in the present setting.

Taken together, these results indicate that AIS-\textit{Sgh} produces thermodynamic estimates that are both internally consistent and in agreement with PT where the two methods concur, while correcting the systematic biases of AIS-\textit{B0} in the low-temperature regime. This holds even for a large-scale trained RBM showcasing the practical utility of AIS-\textit{Sgh}.

\section{Summary and conclusions}
\label{Sec:Conclusions}

In this work, we have presented a systematic study of several fundamental thermodynamic properties and phase transitions of different spin-based systems, including the 2D and 3D Ising model, the 3D EA spin-glass model, and a range of RBMs of increasing complexity. To do that, we have used the Annealed Importance Sampling (AIS) algorithm, finding in all cases good agreement with known critical temperatures and exponents. As a preliminary validation, we have compared the results obtained with AIS against those obtained using standard Metropolis-Hastings and Gibbs Sampling for the Ising model, showing that AIS provides more accurate estimates in the low-temperature regime, where conventional local-update methods tend to become trapped in metastable states and fail to explore the full configuration space.

The central contribution of this work focuses on the validation of the AIS-\textit{Sgh} initialization, already introduced in~\cite{Article_original} as a right tool to detect phase transitions and characterize them when using the Annealed Importance Sampling algorithm of Ref.~\cite{AIS1, AIS2}. Across all RBM systems studied --- from small, exactly solvable Ising, spin-glass and artificial models to large networks trained on real data --- AIS-\textit{Sgh} correctly sampled all thermodynamic phases and reliably detected phase transitions. Crucially, these results were obtained with a deliberately small number of samples and annealing steps, thus demonstrating that AIS-\textit{Sgh} is effective even under stringent computational constraints.

An important unifying aspect of this study is that all systems considered can be mapped onto a bilayer network representation with pairwise interactions. This common formulation allowed the same AIS framework to be applied uniformly across models of very different character, from ordered ferromagnets to fully-connected disordered spin glasses and trained neural-network models. It also enables an efficient parallel implementation of the algorithm, with the bilayer structure being particularly well suited for GPU acceleration, making AIS-\textit{Sgh} a scalable approach for large, high-dimensional systems.

Overall, our work shows that AIS with a proper, yet simple, initialization, is a robust and versatile tool for exploring the thermodynamics of complex spin-like models. The combination of accuracy, phase-awareness, and computational scalability makes this approach especially promising for the analysis of large spin-like energy models and other interacting systems that can be mapped into an RBM, where reliable sampling across multiple phases is essential.

\begin{acknowledgments}
F.M. acknowledges financial support from Ministerio de Ciencia e Innovaci\'on MCIN/AEI/10.13039/501100011033 (Spain) under Grant No. PID2020-113565GB-C21.
E.R. acknowledges financial support from Spanish research grant PID2022-143299OBI00, founded by AEI/10.13039/5011000\textbackslash{}\textbackslash{}11033/FEDER, UE. 
A.P. and J.M. thank financial support provided by the Spanish Ministry of Science, Innovation and Universities. This publication is a part of the I+D+i project with reference PID2024-157478NB-C32, founded by MCIN/AEI/10.13-039/501100011033 and “FEDER Una manera de hacer Europa”. 
F.M. and J.M. thank the {\it Generalitat de Catalunya} for the support through the grant {\it Grup de Recerca SGR-Cat2021 Condensed, Complex and Quantum Matter Group} reference 2021SGR-01411 and to the Polytechnic University of Catalonia-Barcelona Tech through the funding AGRUPS.  
\end{acknowledgments}

\section*{Data Availability}
There are no publicly available research data or software supporting this manuscript. Requests for further information or data should be sent to the authors.

\bibliography{bibliography}

\end{document}

\title{Supplemental Material: Characterization of Phase Transitions with the Annealed Importance Sampling Algorithm}
\author{A. Prat Pou}
\email{arnau.prat.pou@upc.edu}
 \affiliation{Departament de F\'isica, Universitat Polit\`ecnica de Catalunya, Barcelona Tech, Campus Nord B4-B5, E-08034 Barcelona, Spain.}
 
\author{E. Romero}%
 \email{eromero@cs.upc.edu}
\affiliation{Departament de Ci\`encies de la Computaci\'o, Universitat Polit\`ecnica de Catalunya, Barcelona Tech, Campus Nord B4-B5, E-08034, Barcelona, Spain }%

\author{J. Mart\'i}
 \email{jordi.marti@upc.edu}
\affiliation{Departament de F\'isica, Universitat Polit\`ecnica de Catalunya, Barcelona Tech, Campus Nord B4-B5, E-08034 Barcelona, Spain.}%

\author{F. Mazzanti}
 \email{ferran.mazzanti@upc.edu}
\affiliation{Departament de F\'isica, Universitat Polit\`ecnica de Catalunya, Barcelona Tech, Campus Nord B4-B5, E-08034 Barcelona, Spain.}%
\date{\today}
\maketitle

\section{Edwards--Anderson Parameter from AIS Replicas}

The Edwards--Anderson order parameter quantifies the degree of overlap between two independent configurations drawn from two replicas of the system sharing the same Hamiltonian. It is defined as
\begin{equation}
    q_{EA}^{(a,b)} = \frac{1}{N}\sum_i x_i^{(a)} x_i^{(b)},
\end{equation}
where $x_i$ denotes the spin at site $i$, and $a$, $b$ label the two replicas. The thermal average $\langle q_{EA} \rangle$ is then obtained by averaging over many such replica pairs.

In the context of AIS, one typically runs $N_s$ independent chains, each producing a configuration $\mathbf{x}^{(k)}$ together with an importance weight $\omega_k$. To extract reliable statistics, we exploit all $\binom{N_s}{2}$ distinct replica pairs rather than a single pair.

The estimation proceeds in two steps. First, we fix one replica $k$ and estimate its overlap with the remaining $N_s - 1$ replicas using their AIS weights as an importance-sampling measure:
\begin{equation}
    q_{EA}^{(k)} = \frac{\sum_{b \neq k} q_{EA}^{(k,b)}\,\omega_b}{\sum_{b \neq k} \omega_b}.
\end{equation}
This yields $N_s$ replica-specific estimates $\{q_{EA}^{(k)}\}$. Second, we aggregate these estimates by weighting each frozen replica $k$ by its own importance weight $\omega_k$:
\begin{equation}
    \langle q \rangle = \frac{\sum_k q_{EA}^{(k)}\,\omega_k}{\sum_k \omega_k}.
\end{equation}
Substituting the first expression into the second and requiring that each unordered pair $(i,j)$ is counted exactly once, we arrive at the final estimator:
\begin{equation}
    \langle q \rangle = \frac{\displaystyle\sum_{i=1}^{N_s-1}\sum_{j=i+1}^{N_s} q_{EA}^{(i,j)}\,\omega_i\omega_j}{\displaystyle\sum_{i=1}^{N_s-1}\sum_{j=i+1}^{N_s} \omega_i\omega_j}.
\end{equation}
This double sum runs over all $\binom{N_s}{2}$ distinct pairs, ensuring that each pair contributes exactly once and that no replica is correlated with itself.

\section{Partition functions and magnetizations}

\subsection*{Setup and Notation}

Consider an RBM with interaction matrix $\mathbf{W}$, visible bias $\mathbf{b}$, and hidden bias $\mathbf{c}$. The visible and hidden unit vectors are denoted $\mathbf{v}$ and $\mathbf{h}$, respectively, with $N_v$ visible units, $N_h$ hidden units, and a total of $N = N_v + N_h$ units.

The magnetization of a given configuration is defined as
\begin{equation*}
    m(\mathbf{v}, \mathbf{h}) = \frac{1}{N}\left(\sum_i v_i + \sum_j h_j\right),
\end{equation*}
and the energy of that configuration is
\begin{equation*}
    E(\mathbf{v}, \mathbf{h}) = - \mathbf{v}^T\mathbf{W}\mathbf{h} - \mathbf{v}^T\mathbf{b} - \mathbf{h}^T\mathbf{c}.
\end{equation*}
The probability of a state $(\mathbf{v}, \mathbf{h})$ is given by the Boltzmann distribution $p(\mathbf{v}, \mathbf{h}) = e^{-E(\mathbf{v}, \mathbf{h})}/Z$, where
\begin{equation*}
    Z = \sum_{\{\mathbf{v}, \mathbf{h}\}}e^{-E(\mathbf{v}, \mathbf{h})}
\end{equation*}
is the partition function, and the temperature is absorbed into the weights and biases. Throughout, we consider binary units taking values in either $\{0, 1\}^N$ or $\{-1, 1\}^N$, and present results for both cases.

\subsection{Partition Function of an RBM with Binary Units}
\label{subsec:lnZ_RBM}

Starting from the definition of $Z$ and inserting the explicit form of the energy, we have
\begin{equation}
    \label{Eq:Z_def}
    Z = \sum_{\{\mathbf{v}, \mathbf{h}\}} e^{\mathbf{v}^T\mathbf{W}\mathbf{h} + \mathbf{v}^T\mathbf{b} + \mathbf{h}^T\mathbf{c}}.
\end{equation}
Since the exponent is linear in both $\mathbf{v}$ and $\mathbf{h}$, the sum over visible and hidden configurations separates:
\begin{equation}
    Z = \sum_{\{\mathbf{v}\}}e^{\mathbf{v}^T\mathbf{b}} \sum_{\{\mathbf{h}\}}e^{\mathbf{h}^T(\mathbf{v}^T\mathbf{W} + \mathbf{c})}.
\end{equation}
The inner sum over $\mathbf{h}$ further factorises over individual hidden units, since the exponent splits into a sum of independent terms:
\begin{equation}
    \sum_{\{\mathbf{h}\}}e^{\mathbf{h}^T(\mathbf{v}^T\mathbf{W} + \mathbf{c})} =
    \sum_{\{\mathbf{h}\}}\prod_{j=1}^{N_h} e^{h_j(\mathbf{v}^T\mathbf{W}_j + c_j)} =
    \prod_{j=1}^{N_h}\sum_{\{h_j\}} e^{h_j(\mathbf{v}^T\mathbf{W}_j + c_j)},
\end{equation}
where $\mathbf{W}_j$ denotes the $j$-th column of the interaction matrix. Each single-unit sum can now be evaluated analytically. For $h_j \in \{0, 1\}$,
\begin{equation}
    \sum_{\{h_j\}}e^{h_j(\mathbf{v}^T\mathbf{W}_j + c_j)} = 1 + e^{\mathbf{v}^T\mathbf{W}_j + c_j},
\end{equation}
while for $h_j \in \{-1, 1\}$,
\begin{equation}
    \sum_{\{h_j\}}e^{h_j(\mathbf{v}^T\mathbf{W}_j + c_j)} = e^{\mathbf{v}^T\mathbf{W}_j + c_j} + e^{-(\mathbf{v}^T\mathbf{W}_j + c_j)} = 2\cosh{(\mathbf{v}^T\mathbf{W}_j + c_j)}.
\end{equation}
Substituting back, the partition function becomes
\begin{equation}
    \label{Eq:Z_01}
    Z = \sum_{\{\mathbf{v}\}}e^{\mathbf{v}^T\mathbf{b}} \prod_{j=1}^{N_h}\left( 1 + e^{\mathbf{v}^T\mathbf{W}_j + c_j} \right)
\end{equation}
for the $\{0, 1\}$ case, and
\begin{equation}
    \label{Eq:Z_pm1}
    Z = \sum_{\{\mathbf{v}\}}e^{\mathbf{v}^T\mathbf{b}} \prod_{j=1}^{N_h}2\cosh{(\mathbf{v}^T\mathbf{W}_j + c_j)}
\end{equation}
for the $\{-1, 1\}$ case.

\subsection{Mean Magnetization of an RBM with Binary Units}

The thermal average of the magnetization is
\begin{equation}
    \label{Eq:magne_def}
    \langle m \rangle = \frac{1}{Z}\sum_{\{\mathbf{v}, \mathbf{h} \}} m(\mathbf{v}, \mathbf{h})\, e^{\mathbf{v}^T\mathbf{W}\mathbf{h} + \mathbf{v}^T\mathbf{b} + \mathbf{h}^T\mathbf{c}}.
\end{equation}
Substituting the definition of $m$ and separating the sums over visible and hidden units, Eq.~(\ref{Eq:magne_def}) becomes
\begin{equation}
    \label{Eq:magne_LargeSum}
    \langle m \rangle = \frac{1}{ZN} \sum_{\{ \mathbf{v} \} } e^{\mathbf{v}^T\mathbf{b}} \left( \underbrace{\sum_{\{\mathbf{h}\}}\left(\sum_i v_i\right) e^{\mathbf{h}^T(\mathbf{v}^T\mathbf{W} + \mathbf{c})}}_{\text{visible term}}  + \underbrace{\sum_{\{\mathbf{h}\}}\left(\sum_j h_j\right) e^{\mathbf{h}^T(\mathbf{v}^T\mathbf{W} + \mathbf{c})}}_{\text{hidden term}} \right).
\end{equation}

\paragraph{Visible term.}
Since $\sum_i v_i$ does not depend on $\mathbf{h}$, we can factor it out of the sum over hidden configurations. The remaining sum over $\mathbf{h}$ factorises over individual units by the same argument as in Sec.~\ref{subsec:lnZ_RBM}, giving
\begin{equation}
    \label{Eq:magne_sumvis_01}
    \sum_{\{\mathbf{h}\}}\left(\sum_i v_i\right) e^{\mathbf{h}^T(\mathbf{v}^T\mathbf{W} + \mathbf{c})} =
    \left( \sum_{i=1}^{N_v}v_i \right)\prod_{j = 1}^{N_h}\left( 1 + e^{\mathbf{v}^T\mathbf{W}_j + c_j} \right)
\end{equation}
for $\{0, 1\}^N$, and
\begin{equation}
    \label{Eq:magne_sumvis_pm1}
    \sum_{\{\mathbf{h}\}}\left(\sum_i v_i\right) e^{\mathbf{h}^T(\mathbf{v}^T\mathbf{W} + \mathbf{c})} =
    \left( \sum_{i=1}^{N_v}v_i \right)\prod_{j = 1}^{N_h}2\cosh{(\mathbf{v}^T\mathbf{W}_j + c_j)}
\end{equation}
for $\{-1, 1\}^N$.

\paragraph{Hidden term.}
The second sum in Eq.~(\ref{Eq:magne_LargeSum}) is more involved. We split it over individual hidden units and exploit the factorisation of the exponential:
\begin{equation}
    \label{Eq:magne_sumH}
    \begin{split}
    \sum_{\{\mathbf{h}\}}\left(\sum_j h_j\right) e^{\mathbf{h}^T(\mathbf{v}^T\mathbf{W} + \mathbf{c})} &=
    \sum_{k = 1}^{N_h}\sum_{\{\mathbf{h}\}}h_k\, e^{\mathbf{h}^T(\mathbf{v}^T\mathbf{W} + \mathbf{c})} \\
    &= \sum_{k = 1}^{N_h}
    \left(\prod_{j \neq k}\sum_{\{h_j\}} e^{h_j(\mathbf{v}^T\mathbf{W}_j + c_j)}\right)
    \left(\sum_{\{h_k\}} h_k\, e^{h_k(\mathbf{v}^T\mathbf{W}_k + c_k)}\right).
    \end{split}
\end{equation}
Each factor with $j \neq k$ evaluates to $1 + e^{\mathbf{v}^T\mathbf{W}_j + c_j}$ (for $\{0,1\}$), while the $k$-th factor evaluates to $e^{\mathbf{v}^T\mathbf{W}_k + c_k}$, since only the $h_k = 1$ term contributes. Multiplying and dividing each summand by $1 + e^{\mathbf{v}^T\mathbf{W}_k + c_k}$ then yields
\begin{equation}
    \label{Eq:magne_sumhid_01}
    \sum_{k = 1}^{N_h}\sigma\left( \mathbf{v}^T\mathbf{W}_k + c_k \right)\prod_{j = 1}^{N_h}\left( 1 + e^{\mathbf{v}^T\mathbf{W}_j + c_j} \right),
\end{equation}
where $\sigma(x) = 1/(1+e^{-x})$ is the sigmoid function. An analogous computation for $\{-1, 1\}^N$ where the $k$-th factor evaluates to $2\sinh{(\mathbf{v}^T\mathbf{W}_k + c_k)}$ gives
\begin{equation}
    \label{Eq:magne_sumhid_pm1}
    \sum_{k = 1}^{N_h}\tanh{(\mathbf{v}^T\mathbf{W}_k + c_k)}\prod_{j = 1}^{N_h}2\cosh{(\mathbf{v}^T\mathbf{W}_j + c_j)}.
\end{equation}

\paragraph{Final result.}
Adding the visible and hidden contributions, and reinstating the prefactors, we obtain the mean magnetization. For the $\{0, 1\}$ case,
\begin{equation}
    \label{Eq:magne_01}
    \langle m \rangle = \frac{1}{ZN}\sum_{\{\mathbf{v}\}}e^{\mathbf{v}^T\mathbf{b}}\left( \prod_{j = 1}^{N_h} \left( 1 + e^{\mathbf{v}^T \mathbf{W}_j + c_j} \right) \right) \left( \sum_{i=1}^{N_v}v_i + \sum_{j=1}^{N_h}\sigma(\mathbf{v}^T\mathbf{W}_j + c_j) \right),
\end{equation}
and for the $\{-1, 1\}$ case,
\begin{equation}
    \label{Eq:magne_pm1}
    \langle m \rangle = \frac{1}{ZN}\sum_{\{\mathbf{v}\}}e^{\mathbf{v}^T\mathbf{b}}\left( \prod_{j = 1}^{N_h} 2\cosh{(\mathbf{v}^T\mathbf{W}_j + c_j)} \right) \left( \sum_{i=1}^{N_v}v_i + \sum_{j=1}^{N_h}\tanh(\mathbf{v}^T\mathbf{W}_j + c_j) \right).
\end{equation}

Note that the product appearing in each case matches the corresponding partition function integrand from Eqs.~(\ref{Eq:Z_01}) and~(\ref{Eq:Z_pm1}), as expected.

\subsection{Partition Function of an RBM with columns restrictions}

For this derivation, we assume that the biases vanish ($\mathbf{b} = \mathbf{c} = 0$), so that the energy reduces to $E = -\mathbf{v}^T\mathbf{W}\mathbf{h}$. We consider a weight matrix $\mathbf{W}$ with a block-column structure: the $N_h$ hidden units are divided into $k$ groups, where group $s$ contains $M_s$ units (so $\sum_{s=1}^k M_s = N_h$), and all columns within a group are identical. Explicitly, $\mathbf{W}$ takes the form
\begin{equation*}
\mathbf{W} = \left(\begin{array}{ccc|ccc|c|ccc}
a_1^1 & \cdots & a_1^1 & a_1^2 & \cdots & a_1^2 & \cdots & a_1^k & \cdots & a_1^k \\
a_2^1 & \cdots & a_2^1 & a_2^2 & \cdots & a_2^2 & \cdots & a_2^k & \cdots & a_2^k\\
\vdots & \ddots & \vdots & \vdots & \ddots & \vdots & \ddots & \vdots & \ddots & \vdots \\
a_{N_v}^1 & \cdots & a_{N_v}^1 & a_{N_v}^2 & \cdots & a_{N_v}^2 & \cdots & a_{N_v}^k & \cdots & a_{N_v}^k
\end{array}\right),
\end{equation*}
where the $s$-th block contains $M_s$ identical columns.

\subsubsection*{The $\{0,1\}$ Case}

Because all columns in block $s$ are identical, the interaction energy factorises as
\begin{equation}
    \mathbf{v}^T\mathbf{W}\mathbf{h} = \sum_{s=1}^{k}\xi^s \sum_{j \in \text{block } s} h_j, \qquad \xi^s \equiv \sum_{i=1}^{N_v} v_i a_i^s.
\end{equation}
The partition function is therefore
\begin{equation}
    Z = \sum_{\{\mathbf{v}, \mathbf{h}\}} e^{\mathbf{v}^T\mathbf{W}\mathbf{h}} = \sum_{\{\mathbf{v}\}} \prod_{s=1}^{k}\left[\sum_{\{h_j\}_{j \in s}}e^{\xi^s\sum_{j \in s}h_j}\right],
\end{equation}
where the factorisation over blocks follows from the independence of the exponents across groups.

Since the $M_s$ hidden units in block $s$ share the same coupling $\xi^s$, the sum over the $2^{M_s}$ configurations of that block depends only on the total activation $m_s = \sum_{j \in s} h_j \in \{0, 1, \ldots, M_s\}$. Grouping configurations by $m_s$ and counting them with the binomial coefficient,
\begin{equation}
    \sum_{\{h_j\}_{j \in s}}e^{\xi^s\sum_{j \in s} h_j} = \sum_{m_s = 0}^{M_s} \binom{M_s}{m_s} e^{m_s\xi^s}.
\end{equation}
Substituting back and moving the sum over $\mathbf{v}$ inside, the partition function becomes
\begin{equation}
    \begin{split}
    Z &= \sum_{\{\mathbf{v}\}} \prod_{s=1}^k\left[ \sum_{m_s = 0}^{M_s}\binom{M_s}{m_s} e^{m_s\xi^s} \right] \\
    &= \sum_{m_1=0}^{M_1}\cdots\sum_{m_k=0}^{M_k} \binom{M_1}{m_1}\cdots\binom{M_k}{m_k} \sum_{\{\mathbf{v}\}} e^{\sum_{s=1}^k m_s\xi^s}.
\end{split}
\end{equation}
At fixed $\{m_s\}$, the exponent $\sum_s m_s \xi^s = \sum_s m_s \sum_i v_i a_i^s = \sum_i v_i \left(\sum_s m_s a_i^s\right)$ is linear in each $v_i$, so the sum over $\mathbf{v}$ factorises over individual visible units. Evaluating each single-unit sum for $v_i \in \{0,1\}$ gives $\sum_{v_i} e^{v_i x} = 1 + e^x$, and we obtain the final result:
\begin{equation}
    \label{Eq:Z_rows_01}
    Z = \sum_{m_1=0}^{M_1}\cdots\sum_{m_k=0}^{M_k}\; \binom{M_1}{m_1}\cdots\binom{M_k}{m_k} \prod_{i=1}^{N_v}\left(1 + e^{\sum_{s=1}^k m_s a_i^s}\right).
\end{equation}

\subsubsection*{The $\{-1,1\}$ Case}

The argument follows the same structure. We substitute $h_j = 2\tilde{h}_j - 1$ with $\tilde{h}_j \in \{0,1\}$, so that $\sum_{j \in s} h_j = 2m_s - M_s$, where $m_s = \sum_{j \in s}\tilde{h}_j \in \{0,\ldots,M_s\}$ counts the number of $+1$ units in block $s$. The sum over block $s$ then becomes
\begin{equation}
    \sum_{\{h_j\}_{j \in s}}e^{\xi^s\sum_{j \in s} h_j} = \sum_{m_s = 0}^{M_s} \binom{M_s}{m_s} e^{(2m_s - M_s)\xi^s} = e^{-M_s\xi^s}\sum_{m_s = 0}^{M_s}\binom{M_s}{m_s}e^{2m_s\xi^s},
\end{equation}
which, by the binomial theorem, evaluates to
\begin{equation}
    e^{-M_s\xi^s}\left(1 + e^{2\xi^s}\right)^{M_s} = \left(e^{-\xi^s} + e^{\xi^s}\right)^{M_s} = \left[2\cosh(\xi^s)\right]^{M_s}.
\end{equation}
Similarly, the single-unit sum for each visible unit evaluates to $\sum_{v_i \in \{-1,1\}} e^{v_i x} = 2\cosh(x)$. Proceeding as before, the partition function for the $\{-1,1\}$ case is
\begin{equation}
    \label{Eq:Z_rows_pm1}
    Z = \sum_{\{\mathbf{v}\}} \prod_{s=1}^{k}\left[2\cosh(\xi^s)\right]^{M_s} = \sum_{m_1=0}^{M_1}\cdots\sum_{m_k=0}^{M_k}\;\binom{M_1}{m_1}\cdots\binom{M_k}{m_k}\prod_{i=1}^{N_v} 2\cosh\!\left(\sum_{s=1}^k m_s a_i^s\right),
\end{equation}
where in the second equality we have again used $\sum_s m_s \xi^s = \sum_i v_i(\sum_s m_s a_i^s)$ and factorised the visible sum.

Note that Eq.~(\ref{Eq:Z_rows_pm1}) can also be written in the more compact closed form
\begin{equation}
    Z = \sum_{\{\mathbf{v}\}} \prod_{s=1}^{k}\left[2\cosh(\xi^s)\right]^{M_s},
\end{equation}
which makes manifest that each block of $M_s$ identical hidden units contributes a factor of $[2\cosh(\xi^s)]^{M_s}$ to the effective visible-unit weight.

\subsection{Magnetization of an RBM with rows restriction}

We derive the mean magnetization
\begin{equation}
    \langle m \rangle = \frac{1}{ZN}\sum_{\{\mathbf{v}, \mathbf{h}\}}\left(\sum_{i=1}^{N_v} v_i + \sum_{j=1}^{N_h} h_j\right) e^{\mathbf{v}^T\mathbf{W}\mathbf{h}},
\end{equation}
separating visible and hidden contributions:
\begin{equation}
\label{Eq:mag_split}
    \langle m \rangle = \frac{1}{ZN}\left( \underbrace{\sum_{\{\mathbf{v}, \mathbf{h}\}}\left(\sum_{i=1}^{N_v} v_i\right) e^{\mathbf{v}^T\mathbf{W}\mathbf{h}}}_{\text{visible term}} + \underbrace{\sum_{\{\mathbf{v}, \mathbf{h}\}}\left(\sum_{j=1}^{N_h} h_j\right) e^{\mathbf{v}^T\mathbf{W}\mathbf{h}}}_{\text{hidden term}} \right).
\end{equation}

We will treat each term per turn. In this case, we will only present the $\{0,1\}$ case, since, given the symmetry of the problem, the magnetization of the $\{-1,1\}$ will always be equal to zero.

\paragraph{Visible term.}
Summing over hidden configurations exactly as in the partition function derivation and using the block factorisation, the visible term becomes
\begin{equation}
\label{Eq:vis_term}
    \sum_{\{\mathbf{v}, \mathbf{h}\}}\!\left(\sum_i v_i\right) e^{\mathbf{v}^T\mathbf{W}\mathbf{h}}
    = \sum_{\{\mathbf{v}\}}\left(\sum_i v_i\right)\sum_{m_1,\ldots,m_k}\binom{M_1}{m_1}\cdots\binom{M_k}{m_k}
    e^{\sum_i v_i A_i},
\end{equation}
where we have defined $A_i \equiv \sum_s m_s a_i^{(s)}$ for compactness. At fixed $\{m_s\}$, the exponent factorises over visible units. For a single visible unit $v_\ell$, singling it out gives
\begin{equation}
    \left(\sum_{v_\ell \in \{0,1\}} v_\ell\, e^{v_\ell A_\ell}\right)\prod_{i \neq \ell}\left(\sum_{v_i \in \{0,1\}} e^{v_i A_i}\right)
    = e^{A_\ell}\prod_{i \neq \ell}\left(1 + e^{A_i}\right).
\end{equation}
Multiplying and dividing by $1 + e^{A_\ell}$ yields $\sigma(A_\ell)\prod_{i=1}^{N_v}(1 + e^{A_i})$, where $\sigma(x) = e^x/(1+e^x)$ is the sigmoid function. Summing over all $\ell$, the visible term is
\begin{equation}
\label{Eq:vis_final_01}
    \sum_{m_1=0}^{M_1}\cdots\sum_{m_k=0}^{M_k}\binom{M_1}{m_1}\cdots\binom{M_k}{m_k}
    \left(\sum_{i=1}^{N_v}\sigma(A_i)\right)
    \prod_{i=1}^{N_v}\!\left(1 + e^{A_i}\right).
\end{equation}

\paragraph{Hidden term.}
By the exchangeability of units within each block, all $M_s$ terms in block $s$ give identical contributions. Fixing one unit $h_j=1$ in block $s$ and summing the remaining $M_s-1$ units freely,
\begin{equation}
    \sum_{\{h_l\}_{l \in s}} h_j\, e^{\xi^s \sum_{l \in s} h_l}
    = e^{\xi^s}\sum_{m_s=0}^{M_s-1}\binom{M_s-1}{m_s}e^{m_s\xi^s}
    = e^{\xi^s}\left(1 + e^{\xi^s}\right)^{M_s - 1}.
\end{equation}
Using the identity $M_s\binom{M_s-1}{m_s} = (M_s-m_s)\binom{M_s}{m_s}$ to re-index back to limits $0$ to $M_s$, and noting that the shifted exponent $e^{(m_s+1)\xi^s}\prod_{s'\neq s}e^{m_{s'}\xi^{s'}}$ makes the visible sum factorise with effective field $A_i^{(s)} \equiv \sum_{s'} m_{s'} a_i^{(s')} + a_i^{(s)}$, the full hidden term is
\begin{equation}
\label{Eq:hid_final_01}
    \sum_{m_1=0}^{M_1}\cdots\sum_{m_k=0}^{M_k}\binom{M_1}{m_1}\cdots\binom{M_k}{m_k}
    \sum_{s=1}^{k}(M_s - m_s)\prod_{i=1}^{N_v}\!\left(1 + e^{A_i^{(s)}}\right).
\end{equation}

\paragraph{Final result.}
Combining Eqs.~(\ref{Eq:vis_final_01}) and~(\ref{Eq:hid_final_01}),
\begin{multline}
\label{Eq:mag_rows_01}
    \langle m \rangle = \frac{1}{ZN}
    \sum_{m_1=0}^{M_1}\cdots\sum_{m_k=0}^{M_k}
    \binom{M_1}{m_1}\cdots\binom{M_k}{m_k} \\
    \times\left[
    \left(\sum_{i=1}^{N_v}\sigma\!\left(\sum_{s=1}^k m_s a_i^{(s)}\right)\right)
    \prod_{i=1}^{N_v}\!\left(1 + e^{\sum_{s=1}^k m_s a_i^{(s)}}\right)
    \right.\\
    \left.
    +\sum_{s=1}^k(M_s - m_s)
    \prod_{i=1}^{N_v}\!\left(1 + e^{\sum_{s'=1}^k m_{s'} a_i^{(s')} + a_i^{(s)}}\right)
    \right].
\end{multline}